\documentclass[manuscript]{acmart}

\usepackage{multirow}
\usepackage{tabularray}

\usepackage{xcolor}

\definecolor{cbblue}{HTML}{0072B2}
\definecolor{cborange}{HTML}{E69F00}
\definecolor{cbgreen}{HTML}{009E73}

\usepackage[most]{tcolorbox}

\newtcolorbox{Summary}[2][]{title={\bfseries #2},enhanced,
	coltitle=black,
	top=0.17in,
	attach boxed title to top left=
	{xshift=1.5em,yshift=-\tcboxedtitleheight/2},
	boxed title style={size=small,colback=lightgray},#1}

\AtBeginDocument{%
  }

\setcopyright{acmlicensed}
\copyrightyear{2018}
\acmYear{2018}
\acmDOI{XXXXXXX.XXXXXXX}
\acmConference[Conference acronym 'XX]{Make sure to enter the correct
  conference title from your rights confirmation email}{June 03--05,
  2018}{Woodstock, NY}
\acmISBN{978-1-4503-XXXX-X/2018/06}

\begin{document}

\title{LabelMate: An LLM-Driven Framework for Refined Issue Report Labeling}


\author{Liam Johnston}
\affiliation{
\institution{Queen's University}
\city{Kingston}
\state{Ontario}
\country{Canada}
}
\email{24rrvk@queensu.ca}

\author{Shayan Noei}
\affiliation{
\institution{Queen's University}
\city{Kingston}
\state{Ontario}
\country{Canada}
}
\email{s.noei@queensu.ca}

\author{Maram Assi}
\affiliation{
\institution{Universit\'e du Qu\'ebec \`a Montr\'eal}
\city{Montr\'eal}
\state{Quebec}
\country{Canada}
}
\email{assi.maram@uqam.ca}

\author{Ying Zou}
\affiliation{
\institution{Queen's University}
\city{Kingston}
\state{Ontario}
\country{Canada}
}
\email{ying.zou@queensu.ca}

\renewcommand{\shortauthors}{Johnston et al.}

\begin{abstract}

Software users often submit issue reports to a product’s issue tracking system to report defects, suggest enhancements, or raise other product-related concerns. Labeling these issue reports supports effective planning % , enhances task discoverability, 
and improves community engagement. However, many issue reports remain unlabeled due to the substantial manual effort required to design an appropriate label taxonomy, then assign suitable labels from this taxonomy to new issue reports. Existing automated labeling approaches attempt to mitigate these challenges. However, they suffer from key limitations, such as extensive manual intervention, the assignment of generic labels, and a dependence on existing labeled datasets. To address these limitations, we propose \textit{LabelMate}, a novel Large Language Model (LLM)-driven framework that (1) derives a comprehensive, project-specific label set from historical issue reports and (2) automatically assigns relevant labels to new issue reports without requiring any pre-labeled training data. We evaluate \textit{LabelMate} on 16,500 issue reports from 30 popular and diverse GitHub repositories. Based on this dataset, our approach generates a coherent list of 275 labels and achieves an average labeling accuracy of 89.84\%, a statistically significant improvement over existing generic label assigning approaches. These results demonstrate that \textit{LabelMate} offers an efficient, domain-adaptive solution to streamline the issue labeling process.

% by enabling contributors to efficiently identify issues relevant to their expertise.

% for each project

%\maram{did we compare to existin benchmarks? I don't think so...} \liam{Three things: (1) I think its important to say based on this dataset because the approach will generate a different label list and achieve different accuracy on different datasets or should we assume the reader knows this, (2) we compare the performance of our approach against the 4-label approach and the original labels assigned to the issue reports (see Table \ref{tab:RAG_results}) so I think it should be appropriate to say this but we can discuss, and (3) should we not say that it is significant according to statistics?? or is it implied when you say significant that it is in terms of statistics??} 

\end{abstract}

\begin{CCSXML}
<ccs2012>
   <concept>
       <concept_id>10011007.10011074.10011111.10011696</concept_id>
       <concept_desc>Software and its engineering~Maintaining software</concept_desc>
       <concept_significance>500</concept_significance>
       </concept>
   <concept>
       <concept_id>10010147.10010178.10010179.10003352</concept_id>
       <concept_desc>Computing methodologies~Information extraction</concept_desc>
       <concept_significance>300</concept_significance>
       </concept>
 </ccs2012>
\end{CCSXML}

\ccsdesc[500]{Software and its engineering~Maintaining software}
\ccsdesc[300]{Computing methodologies~Information extraction}

\keywords{Large Language Models (LLMs), issue reports, labeling}

\received{20 February 2007}
\received[revised]{12 March 2009}
\received[accepted]{5 June 2009}

\maketitle

\section{Introduction}
\label{sec:intro}

Software users communicate defects, enhancement requests, and other product-related concerns through issue reports submitted to the project’s issue tracking system \cite{kim2021empirical}. An issue report typically comprises a concise title summarizing the user's concern and a detailed body that provides additional context which enables project contributors to understand and address the reported issue \cite{montgomery2022alternative}. Effective triaging and management of issue reports is crucial to a product's long-term success as it directly affects software quality, reputation, and contributor engagement \cite{junior2021label, zanetti2013categorizing}. % For instance, assigning issue reports to contributors whose expertise aligns with the required resolution steps can accelerate issue resolution \cite{garzarelli2008open}. 
This process can be enhanced through the use of \textit{labels}, which are short descriptive tags that categorize issue reports into functional and technical dimensions (e.g., \textit{feature request}, \textit{user experience}, \textit{bug}, \textit{performance}, \textit{security}). % , as in platforms like GitHub, contributors can query issue reports by label, enabling them to quickly identify tasks that match their expertise \cite{santos2021can}. %This label-based filtering improves efficiency in large repositories with a high number of such reports . 
Studies have shown that labeled issue reports attract significantly more attention (e.g., subscribers, assignments, and comments) than those without labels \cite{junior2021label} and that GitHub repositories that label their issue reports resolve them more efficiently than those who do not \cite{kim2021empirical}.

Despite the benefits of labeling, a large proportion of issue reports remain unlabeled in practice. J{\'u}nior et al.~\cite{junior2021label} find that only 46.07\% of over 10 million issue reports across more than 13,000 GitHub repositories were labeled. Similarly, Kim and Lee \cite{kim2021empirical} report that only 54.59\% of more than 13 million issue reports from over 14,000 GitHub repositories were labeled. These findings highlight a persistent gap between the recognized value of labeling and its limited adoption in real-world projects. Two main challenges contribute to this limited adoption. First, deriving a label list that is suitable for a given project can be a significant barrier to entry as it requires understanding the project's structure and workflows as well as anticipating the types of issues that may arise \cite{junior2021label}. Second, even after a label list has been defined, assigning relevant label(s) to each incoming issue report is often a manual, labour-intensive, and time-consuming process \cite{fan2017road}. 

Several approaches have been proposed to alleviate these adoption challenges by simplifying label creation and automating label assignment. However, each approach comes with one of the following notable limitations: 

\begin{itemize}
    \item \textbf{Existing Issue Report Category Extraction is Rigid and Labour-Intensive.} Embedded Topic Modeling (ETM) \cite{dieng2020topic} has been employed to identify latent categories within issue reports and the resulting topics can subsequently be interpreted as candidate labels for issue reports. While promising, this approach requires the number of labels to be specified a priori and relies heavily on manual inspection and interpretation of the model outputs to determine suitable labels.
    
    \item \textbf{Assignment of Generic Labels.} Several existing automated labeling approaches restrict labels to a set of two to four broad categories (e.g., \textit{bug}, \textit{feature}, \textit{question}, \textit{documentation}) \cite{fan2017road, aracena2024applying, colavito2024leveraging, kallis2021predicting, heo2024comparison}. This oversimplified taxonomy does not reflect the more fine-grained labeling practices commonly adopted in real-world software projects. For example, the median number of labels used across the 30 GitHub repositories in our issue report dataset is 124, which demonstrates that in practice, projects employ more detailed labeling schemes than the set of 4 labels used in these works. For a specific example, the GitHub repository \textit{golang/go}\footnote{\href{https://github.com/golang/go}{https://github.com/golang/go}} uses labels such as \textit{Security}, \textit{compiler/runtime}, \textit{Performance}, and \textit{Refactoring}. These details are not captured by the four-label taxonomy consisting only of the labels \textit{bug}, \textit{feature}, \textit{question}, and \textit{documentation}, which highlights that the four-label taxonomy fails to capture the richness and specificity of labels used in practice.
    
    \item \textbf{Dependence on Existing Labeled Datasets.} A more advanced existing automated labeling approach attempts to assign more granular labels used in practice by training models on labeled issue reports from real-world projects \cite{heo2024comparison}. However, this approach requires the existence of a consistent and sufficiently large labeled dataset within each project. As a result, this method does not effectively mitigate the adoption challenges.

    \item \textbf{Unconstrained LLMs Produce Noisy, Unusable Label Spaces.} In our own motivational study, we examine the performance of LLMs in labeling issue reports without constraining the label space. Our findings show that under these conditions, LLMs generate an excessive number of labels, many of which are overlapping or overly specific. For example, when applied to a set of 13,210 issue reports, the LLM \textit{Llama-3.1-8B-Instruct} produced 8,020 unique labels, of which 61.77\% were generated from a single issue report. If adopted in practice, such inflated label sets would hinder navigation and retrieval of related issues, thereby reducing the utility of labeling in software maintenance and project management. 

\end{itemize}

Taken together, prior work demonstrates the potential of automated label derivation and assignment but highlights the need for more practical solutions.

To overcome the limitations of existing labeling approaches, we introduce \textit{LabelMate}, a framework that leverages the advanced natural language understanding of Large Language Models (LLMs). \textit{LabelMate} alleviates the adoption challenge of deriving a project-appropriate label list by offering a flexible, low-effort framework that improves on both extremes of prior work: it is less rigid and labour-intensive than ETM-based taxonomy construction and it produces far more practical label lists than the overly generic four-category sets used in earlier studies. The framework achieves this through the refinement of candidate labels generated by LLMs from historical issue reports into a coherent, project-relevant taxonomy. It also employs an LLM for the label assignment task, thereby automating this process and mitigating the substantial time investment typically required for manual labeling. In addition, we introduce a retrieval-augmented generation (RAG)-based method that dynamically narrows the label list provided to the LLM by retrieving semantically similar, previously labeled issues. This targeted conditioning improves label assignment accuracy by preventing the model from selecting irrelevant labels. We also include an approach to automatically derive a labeled dataset, removing the need for projects to supply one in advance. 

This design enables practical adoption for collaborative software projects with minimal setup. Teams with existing taxonomies can immediately use our RAG-based assignment method to maintain consistent labeling aligned with their current scheme. Teams seeking to improve their taxonomy or create a new one may either adopt our curated list of 275 labels from a set of 30 diverse GitHub repositories or generate their own project-specific taxonomy using our taxonomy induction pipeline. 

We evaluate \textit{LabelMate} using 16,500 issue reports from 30 popular and diverse GitHub repositories using multiple open-source LLMs for both taxonomy induction and label assignment. This evaluation is guided by the four research questions (RQs) outlined below:
% \maram{you are adding some unecessary details in the RQ descriptions. It should provide 1 sentence motivation, brief high level approacha and outcome.} \liam{Is this more what you're looking for?}

\noindent \textbf{RQ1: How can we derive a coherent label list for a set of issue reports?} 

\vspace{0.5em}

\hspace*{1.5em}\parbox{\dimexpr\linewidth-2.5em}{
Given the findings from our motivational study, we propose that LLMs would benefit from being restricted to assigning labels from a pre-defined list. However, previous approaches limit labels to either overly generic sets or require an existing set of labels. As a result, we propose a method for constructing a label list based on a set of historic issue reports. This involves taking the labels generated by LLMs from the historic issue reports in an unconstrained setting (i.e., our motivational study) and consolidating them, through methods such as clustering synonymous labels based on the similarity of their embeddings, to obtain a list of meaningful labels free of redundancy. Using this method on our dataset, we produce a set of 275 distinct and descriptive labels to assign to issue reports.
}

\vspace{0.5em}

\noindent \textbf{RQ2: How do LLMs assign labels to issue reports from our coherent label list?}

\vspace{0.5em}

\hspace*{1.5em}\parbox{\dimexpr\linewidth-2.5em}{
After deriving our coherent list of 275 labels in the second research question, we highlight its utility by showing that the labels assigned by LLMs to issue reports from this list are more semantically aligned with the contents of the issue report (measured via cosine similarity) compared to existing approaches. Specifically, the labels assigned by the LLM \textit{Qwen2.5-7B-Instruct} from our derived list achieved the highest average cosine similarity with the contents of the issue report (with a value of 0.178) across all tested combinations of label assigners and label lists.
}

\vspace{9em}

\noindent \textbf{RQ3: Can we enhance our labeling pipeline using retrieval-augmented generation?}

\vspace{0.5em}

\hspace*{1.5em}\parbox{\dimexpr\linewidth-2.5em}{
Providing the entire list of 275 labels in the prompt to LLMs can cause issues such as increasing inference cost and the risk of hallucinated label assignments given that many labels in the list may be irrelevant to certain issue reports. To address these limitations, we assess a RAG-based approach that retrieves a context-specific subset of the label list for each issue report. This approach achieves a higher overall label accuracy (89.84\%) and reduces prompt token usage and runtime compared to including the entire list of 275 labels in the prompt, thereby offering both performance and efficiency gains.
}

\vspace{0.5em}

\noindent \textbf{RQ4: How does our coherent label list align with label lists of existing collaborative software repositories?}

\vspace{0.5em}

\hspace*{1.5em}\parbox{\dimexpr\linewidth-2.5em}{
To assess the practical relevance of our derived label taxonomy, we compare our derived list of 275 labels with the label lists used by existing collaborative software repositories. Specifically, we examine the degree of coverage between labels in our derived list with labels in the existing taxonomies of four GitHub repositories of varying popularity that were not included in our dataset used to construct the label list. This involves computing how many of the labels in the existing taxonomy are covered by labels in our label list at various embedding-based similarity thresholds. Our results show that our list provides substantial coverage of up to 100\% at thresholds as high as 0.4, indicating that the labels produced by our approach reflect widely used issue concepts and provide strong coverage of real-world labeling practices.
}

\vspace{1em}

The following are the major contributions of our work:

\begin{itemize}
    \item \textit{LabelMate}, a novel end-to-end, domain-adaptive LLM-driven issue report labeling framework that enables (1) the derivation of a comprehensive and customized list of labels for any project based on historical issue reports and (2) the automated assignment of these labels to incoming issue reports.

    \item A replication package\footnote{Replication package available at  \href{https://github.com/24rrvk/LLMIssueLabeling}{https://github.com/24rrvk/LLMIssueLabeling}} containing (1) the implementation of our approach, including LLM prompts and usage instructions, and (2) the results we obtained from running our approach, including datasets, the derived label list, and labels assigned to issue reports using the various methods and (3) an example of how our automated labeling pipeline can be implemented in real-world issue triaging workflows.

\end{itemize}

\textbf{Paper Organization.} The remainder of this paper is structured as follows: Section \ref{sec:experimental_setup} details the overall approach, the models employed in our framework, the dataset curation process, and the data pre-processing pipeline. Section \ref{sec:motivational_study} describes the motivation, approach, and findings of our two motivational studies. Section \ref{sec:rqs} describes the motivation, approaches, and findings of our four research questions. Section \ref{sec:implications} discusses implications of this study. Section \ref{sec:threats} lists possible threats to validity. Section \ref{sec:related_work} positions our work within the broader literature on automated issue report labeling and using LLMs to label, annotate, or categorize other software engineering content. Section \ref{sec:conclusion} concludes the study.

% \section{Approach}
% \label{sec:approach}
% \input{sections/approach}

\section{Experimental Setup}
\label{sec:experimental_setup}

This section details the overall approach, the models employed in our framework, the dataset curation process, and the data pre-processing pipeline.

\subsection{Overview of Approach}

\begin{figure*}
    \centering\includegraphics[width=1.0\linewidth]{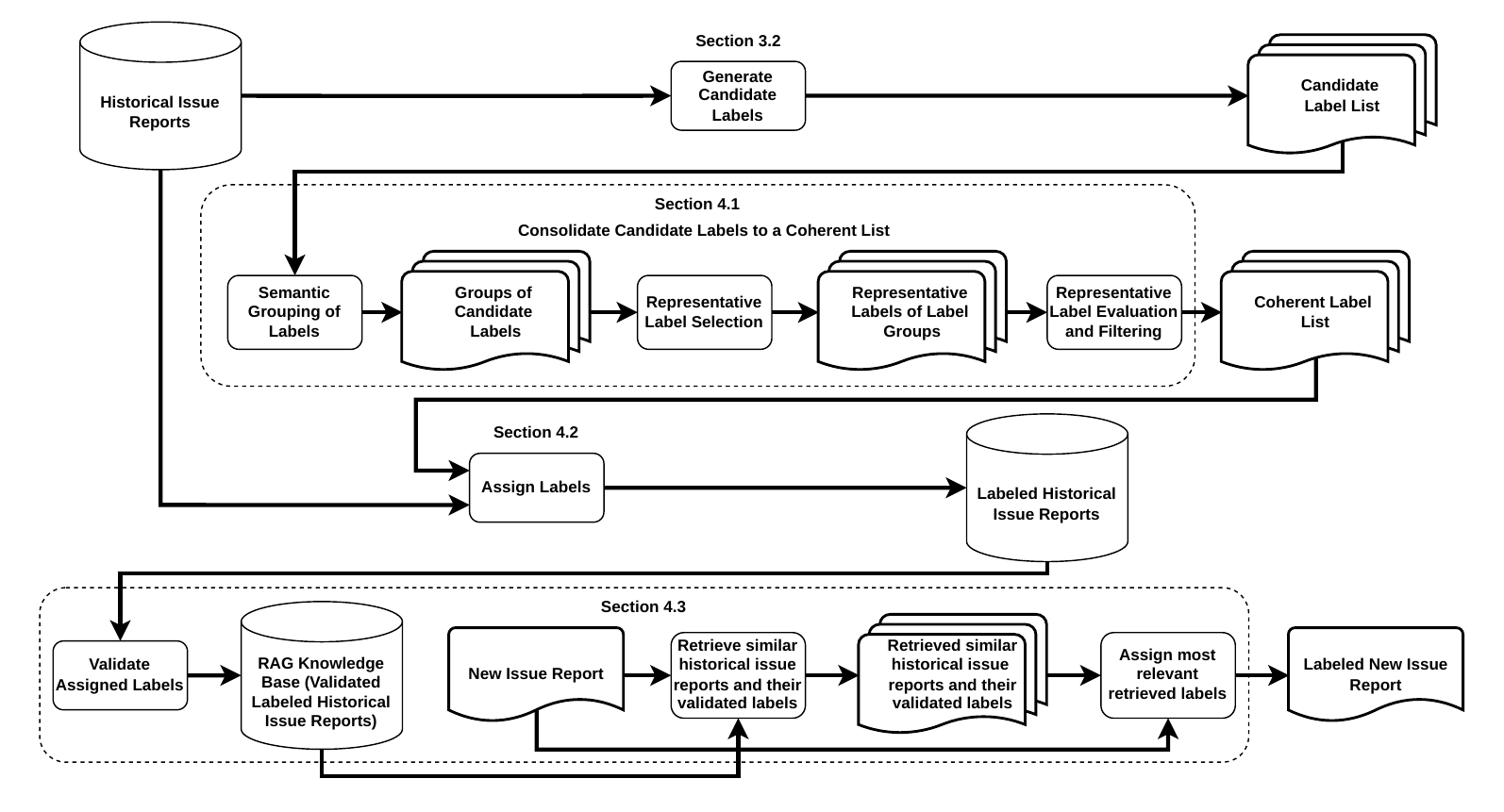}
    \caption{Outline of Overall Approach.}
    \label{fig:approach}
\end{figure*}

\textit{LabelMate} is a framework designed for automated and domain-adaptive labeling of software issue reports using large language models (LLMs). The overall architecture is illustrated in Figure~\ref{fig:approach}. To build a coherent label list, we first prompt LLMs to generate candidate labels from historical issue reports. These candidates are then refined through semantic clustering to merge synonymous terms and eliminate redundancies and representative label selection and evaluation, resulting in a coherent label taxonomy that captures the key themes within historical issues. The framework then employs LLMs to assign labels from the generated taxonomy to new issue reports. To enhance labeling accuracy, this stage adopts a context-specific labeling strategy that retrieves the most relevant subset of labels from the taxonomy using a retrieval-augmented generation (RAG) mechanism before performing label assignment.

\subsection{Model Selection}
\label{sec:models_selected}
\textit{LabelMate} incorporates three key model components: (1) \textbf{Label Assigner LLMs}, which are used to generate the candidate labels that are refined to derive our label list and assign labels from the generated list to new issue reports, (2) a \textbf{Label Evaluator LLM}, which is used to evaluate the accuracy of label assignments \cite{thakur2025judging, nahum2025llms, zhou2025se}, and (3) a \textbf{Text Embedding Model} which encodes both labels and issue reports for semantic clustering and retrieval. Below, we describe and justify the selection of LLMs and text embedding model used in our framework.

\subsubsection{Label Assigner LLMs}
\label{sec:label_assigner_LLMs} 

We employ three instruction-tuned, open-source LLMs, i.e., \textit{gemma-2-9b-it}\footnote{\href{https://huggingface.co/google/gemma-2-9b-it}{https://huggingface.co/google/gemma-2-9b-it}}, \textit{Llama-3.1-8B-Instruct}\footnote{\href{https://huggingface.co/meta-llama/Llama-3.1-8B-Instruct}{https://huggingface.co/meta-llama/Llama-3.1-8B-Instruct}}, and \textit{Qwen2.5-7B-Instruct}\footnote{\href{https://huggingface.co/Qwen/Qwen2.5-7B-Instruct}{https://huggingface.co/Qwen/Qwen2.5-7B-Instruct}}, as our label assigner models. They are referred to in the remainder of the paper as \textit{Gemma}, \textit{Llama}, and \textit{Qwen} respectively. These models are chosen for three key reasons. First, their instruction tuning enables robust adherence to prompt directives, which is essential for consistent label generation~\cite{ouyang2022training}. Second, being open-source, they ensure transparency, reproducibility, and accessibility for academic and industrial use. Third, their moderate parameter size (7-9B) provides an effective trade-off between performance and computational cost~\cite{jiang2025aixcoder}, facilitating efficient deployment of \textit{LabelMate} without requiring extensive GPU resources.

\subsubsection{Label Evaluator LLM} 

We select \textit{deepseek-r1:70b}\footnote{\href{https://ollama.com/library/deepseek-r1:70b}{https://ollama.com/library/deepseek-r1:70b}} as our label evaluator LLM due to its advanced reasoning capabilities. This allows us to determine \textit{why} the model judged a label as accurate or not for a given report \cite{guo2025deepseek}. Although this model is substantially larger and therefore less accessible than the assigner models, i.e., the label evaluator LLM has 70 billion parameters whereas the label assigner LLMs range from having 7 to 9 billion parameters, it is only required to build the labeled knowledge base used in our RAG framework. Consequently, users can benefit from the deeper reasoning of a large model without needing to deploy it repeatedly, preserving the overall efficiency and accessibility of our approach. Furthermore, as with the label assigner LLMs, this model is open-source, which upholds \textit{LabelMate's} ease of adoption.

\subsubsection{Text Embedding Model}
\label{sec:text_embedding_models_selected} 

We adopt \textit{all-mpnet-base-v2}\footnote{\href{https://huggingface.co/sentence-transformers/all-mpnet-base-v2}{https://huggingface.co/sentence-transformers/all-mpnet-base-v2}} as our text embedding model as it is considered to be state-of-the-art and is designed to capture deep semantic relationships. It is a variant of the pretrained \textit{mpnet-base} model \cite{NEURIPS2020_c3a690be} that was fine-tuned on 1B sentence pairs to enhance performance on semantic similarity tasks. This fine-tuning makes it well-suited for measuring semantic alignment and supporting retrieval in RAG-based pipelines and as a result, it has been previously employed in such contexts \cite{cheng2023neural, xureasoning, colangelo2025comparative}. Additionally, as with the selected LLMs, it is open-source to maintain \textit{LabelMate's} ease of adoption.

\subsection{Dataset Curation}
\label{sec:dataset_curation}

To ensure representativeness of real-world software engineering practices, we select repositories that are popular, actively maintained, and whose issue reports correspond to concrete software engineering problems resolved through code changes. We curate our dataset using GitHub’s REST API\footnote{\url{https://docs.github.com/en/rest?apiVersion=2022-11-28}} and apply the following inclusion criteria: popularity, activity and completeness, and relevance.

First, we restrict our selection to repositories ranked among the 500 most-starred on GitHub, as this metric reflects both project maturity and sustained community engagement~\cite{borges2016predicting}. Second, to ensure the inclusion of meaningful, actionable issue reports, we require each repository to contain at least 100 closed issues that were opened after 2023 and associated with at least one resolution patch. Third, we focus exclusively on issue reports related to software engineering activities such as bug fixes, feature requests, and performance optimizations, excluding repositories whose issues were unrelated to software development.

We conducted the collection process on March 25, 2025, compiling a total of 16,500 issue reports from 30 GitHub repositories. To emulate a realistic project scenario, we partition the dataset into training and test sets. The training set represents historical issue reports used to construct the % project-specific 
label list and RAG knowledge base, while the test set represents newly submitted issue reports used for label assignment. For each repository, the 20\% most recently opened issue reports were allocated to the test set, while the remaining 80\% formed the training set. We adopt the 80\%, 20\% train-test split as it is considered to be common practice \cite{GunKurnia_2024}. This resulted in a total of 13,210 issue reports in the train set and 3,290 issue reports in the test set.

% Furthermore, to ensure~\shayan{explain the reason for splitting, for example, avoiding overfitting, or make sure our approach has never seen the data we are testing}, we split the issue reports from each repository into 80\% training and 20\% testing sets~\shayan{cite}. This split was conducted based on the creation date of the issue reports: for each project, we assign 20\% most recently opened to the test set while the remaining 80\% formed the training set. This temporal split mimics real-world conditions: training on past data to predict future outcomes~\shayan{cite}. \sout{This resulted in a total of 13,210 issue reports in the train set and 3,290 issue reports in the test set.}~\shayan{move sum to above where I mentioned}

% We employ the training set as the set of ``historical issue reports'' used to generate candidate labels for the label list and to serve as our RAG database. We employ the test set as the set of ``new issue reports'' used to evaluate the performance of our prompting strategies in the \textit{Automated Assignment of Labels} stage of our approach. 

% \liam{I don't think we need this since we explain this in the approach section}

\subsection{Data Pre-processing}
\label{sec:issue_report_pre-processing_steps}

\begin{figure*}
        \centering
    \includegraphics[width=1.0\linewidth]{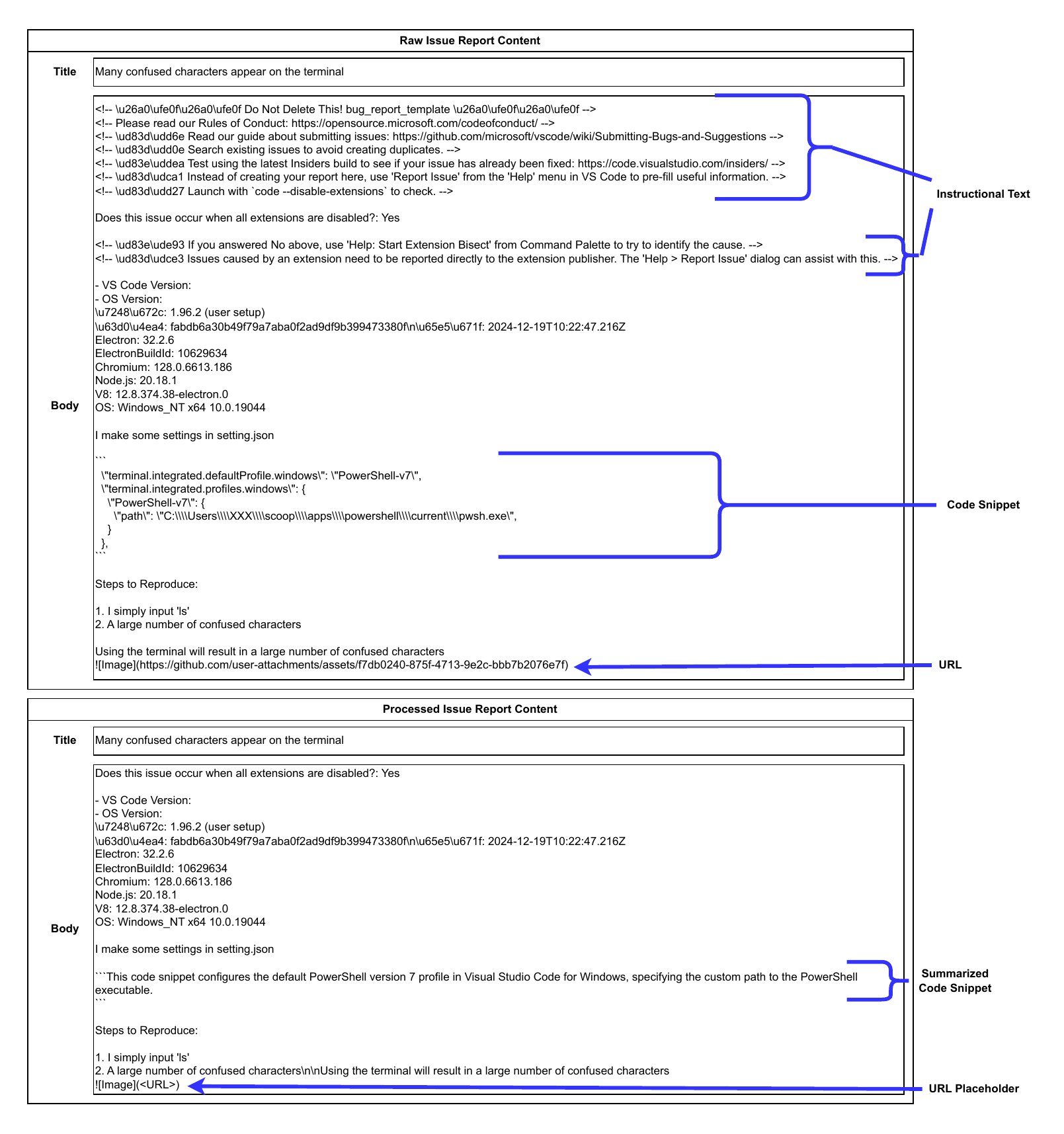}
    \caption{The raw content and content after applying our pre-processing steps of issue \#236994 of the GitHub repository \textit{microsoft/vscode}. The summary of the code snippet was generated by \textit{Qwen2.5-7B-Instruct}.
    }
    \label{fig:issue_report_preprocessing_example}
\end{figure*}

Issue reports on platforms such as GitHub typically contain a mixture of natural-language descriptions and technical artifacts. Besides the title and main body text, users frequently include items such as URLs, which sometimes point to images, code snippets, shell scripts, and output logs. Many projects also use issue templates containing instructional text. Such text is enclosed within HTML comments in the raw data but is hidden in the rendered web interface \cite{githubQuickstartWriting} and thus not visible to users browsing the issue. Figure \ref{fig:issue_report_preprocessing_example} shows an example of an issue report in our dataset which includes HTML comment blocks from the repository's bug report template, a JSON configuration code snippet, and an image embedded as a URL.

To prepare the issue reports for use as input to our LLMs, we apply the pre-processing steps described below, following established practices in software repository mining and natural language processing for issue report datasets~\cite{jokhio2021mining, subramanian2013making}.

\begin{enumerate}
    \item \textbf{Remove null characters:} We first strip all null characters (``\textbackslash0'') from the text as they can interfere with text parsing and string processing \cite{ssojetNull0}.
    \item \textbf{Standardize URLs:} All URLs are replaced with the placeholder token ``\textless URL\textgreater'' to ensure that links do not introduce irrelevant lexical variation ~\cite{mediumFromText}.
    \item \textbf{Remove HTML comment blocks:} As mentioned, many projects include instructional text in their issue templates which is enclosed within HTML comments (e.g., ``\textless{}!- - ... - -\textgreater{}''). This content is not rendered in the GitHub web interface because it is not part of the user's actual problem description and is therefore irrelevant to the issue resolver. For the same reason, it is unnecessary to provide this content to the the LLMs for their tasks in this work, so we remove it during pre-processing.
    \item \textbf{Summarize technical artifacts:} Issue report bodies frequently contain code snippets, shell scripts, and output logs. Instead of requiring the LLM to interpret these raw artifacts directly, we first prompt it to generate a concise natural-language summary of each snippet. We then replace the original snippet with its summary in the processed issue report. This reduces input noise while preserving the semantic information required for label assignment \cite{mastropaolo2024towards}. The prompts used to summarize code snippets, shell scripts, and output logs can be found in the replication package\footnote{\href{https://github.com/24rrvk/LLMIssueLabeling/tree/main/Results\_and\_Prompts/prompts/technical\_summarization\_prompts}{https://github.com/24rrvk/LLMIssueLabeling/tree/main/Results\_and\_Prompts/prompts/technical\_summarization\_prompts}}.

\end{enumerate}

Figure \ref{fig:issue_report_preprocessing_example} illustrates an example of applying these steps to the raw content of an issue report in our dataset. It highlights the conversion of a URL to the placeholder token, the removal of all instructional text enclosed within the HTML comment tags, and the replacement of the code snippet with an LLM-generated summary.

% \subsection{Research Methods}
% ~\shayan{we may need this subsection to discuss our formulas, statistical tests, etc, I'll leave it blank maybe we move some from RQs here}

\section{Motivational Studies}
\label{sec:motivational_study}

This section describes the motivation, approach, and findings of our two motivational studies (MSs).

\subsection{MS1: How do LLM assignments of labels to issue reports compare to the original labels assigned to these issue reports when assigning from the same set of labels?}
\label{sec:llms_assigning_original_labels}

\subsubsection{Motivation}

Previous studies using LLMs to assign labels to issue reports have restricted the label space to a set of two to four broad categories (e.g., \textit{bug}, \textit{feature}, \textit{question}, \textit{documentation}) \cite{aracena2024applying, colavito2024leveraging}. However, as discussed in Section \ref{sec:intro}, real-world collaborative software repositories often employ more fine-grained labels. This mismatch limits the ecological validity of prior evaluations. As a result, in this study, we examine how LLMs assign labels drawn directly from the label taxonomies of the actual repositories and compare these assignments against the labels originally assigned to these issue reports. To the best of our knowledge, this is the first work to investigate LLM performance under these realistic labeling conditions.

\subsubsection{Approach}

The first step in the approach of this MS is to obtain the label list of each of the 30 GitHub repositories in our dataset. We then remove all labels that refer to the \textit{status of the issue's resolution} (e.g., \textit{help-wanted}, \textit{wontfix}, \textit{needs reproduction}) from each repository's label list since the status of an issue report changes over time. Consequently, assigning such labels based solely on the issue report text is not meaningful for LLM-based labeling. The classification of labels as either referring to the status of an issue's resolution or not can be viewed in our replication package\footnote{\url{https://github.com/24rrvk/LLMIssueLabeling/tree/main/Results_and_Prompts/original_label_lists/projects_in_dataset}}.

\begin{figure*}
    \centering
    \includegraphics[width=0.8\linewidth]{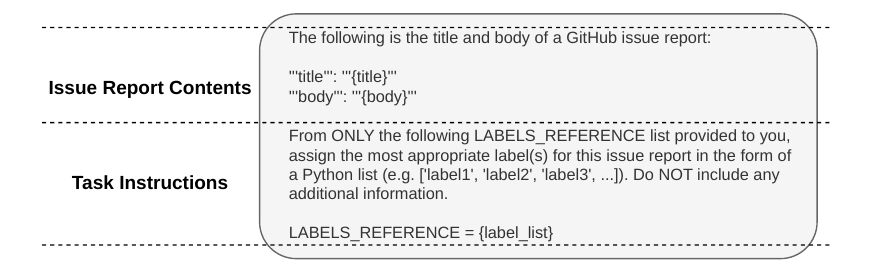}
    \caption{Full Label List and RAG Labels only prompt.
    }
    \label{fig:full_label_list_and_RAG_labels_only_prompt}
\end{figure*}

Next, we experiment with each of our three label assigner LLMs assigning labels to the 3,290 issue reports in our test set using three different prompts. The first is the \textit{Full label list} prompt, which is shown in Figure \ref{fig:full_label_list_and_RAG_labels_only_prompt}. Specifically, we insert the label list of the repository of the issue report to label, with labels referring to the status of the issue's resolution removed, in the \{label\_list\} placeholder.

We further experiment with two retrieval-augmented generation (RAG)-based prompts. In order to set up a RAG-based system, the first step is to construct a RAG knowledge base, which is a database that allows the model to utilize external, domain-specific information during the generation process \cite{gao2023retrieval}. Our RAG knowledge base consists of the original label assignments to the 13,210 issue reports in the training set, with labels referring to the status of the issue's resolution removed. We embed these issue reports using our text embedding model (\textit{all-mpnet-base-v2}) and store these embeddings using the \textit{Faiss} library \cite{douze2024faiss} as it has been widely adopted in previous works in semantic-search and retrieval-based systems \cite{sunho2023duplicate, xian2024bert, li2025enhancing}. In our case, it enables efficient similarity-based retrieval of contextually relevant historical issue reports and their original assigned labels from our knowledge base.

\begin{figure*}
    \centering
    \includegraphics[width=0.8\linewidth]{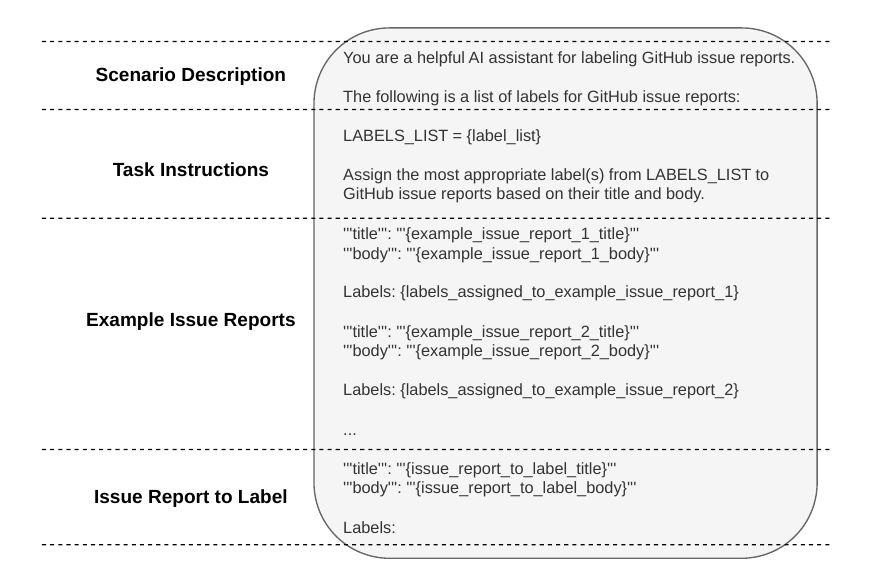}
    \caption{RAG labels and issue reports prompt.}
    \label{fig:RAG_labels_and_issue_reports_prompt}
\end{figure*}

To assign labels to an incoming issue report using our RAG-based system, the first step is to embed the incoming issue report's title and body to represent their semantic meaning \cite{chersoni2021decoding}. We then retrieve the $k$ issue reports in the RAG knowledge base from the same repository as the incoming issue report with the most similar embeddings, and therefore most similar semantic meaning, using the \textit{Faiss} library \cite{douze2024faiss}, along with the original labels assigned to them. The LLM is then prompted to assign the most relevant labels of those retrieved to the incoming issue report. We experiment with two different prompts using the retrieved information:

\begin{itemize}
    \item \textbf{Labels Only:} Prompt shown in Figure \ref{fig:full_label_list_and_RAG_labels_only_prompt} where we insert the retrieved labels into the \{label\_list\} placeholder.
    
    \item \textbf{Labels and Issue Reports:} Prompt shown in Figure \ref{fig:RAG_labels_and_issue_reports_prompt} that includes not only the retrieved labels in the \{label\_list\} placeholder, but also the retrieved issue reports which the retrieved labels were assigned to.
     
\end{itemize}

We experiment with the two RAG-based prompts across $k = 1$ to $k = 19$.

\textbf{Evaluation Metrics.} We compare the alignment of labels assigned by LLMs to issue reports using each prompt with the labels originally assigned to the issue reports using the standard metrics \textit{precision}, \textit{recall}, and \textit{F1 score}:

\begin{equation}
    \text{Precision} = \frac{TP}{TP + FP}
\end{equation}

\begin{equation}
    \text{Recall} = \frac{TP}{TP + FN}
\end{equation}

\begin{equation}
    \text{F1 Score} = \frac{2 \times \text{Precision} \times \text{Recall}}{\text{Precision} + \text{Recall}} = \frac{2TP}{2TP + FP + FN}
\end{equation}

where 

\begin{itemize}
    \item \textit{TP}, or \textit{True Positives}, are when the LLM assigned a label to an issue report that was originally assigned to the issue report.
    \item \textit{FP}, or \textit{False Positives}, are when the LLM assigned a label to an issue report that was not originally assigned to the issue report.
    \item \textit{FN}, or \textit{False Negatives}, are when the LLM did not assign a label to an issue report that was originally assigned to the issue report.
\end{itemize}

We further compute the average number of labels assigned to the issue reports by each LLM using each prompt, as well as the average runtime and average number of tokens. Note that runtime excludes model and data loading, and only measures the time required to construct the prompt using the loaded data and obtain the LLM's response. For the RAG-based prompts, this includes retrieving information from the RAG database, but does not include the time required to load the database. Additionally, the number of tokens for the same prompt and issue report may vary across LLMs for two reasons: (1) each LLM uses its own tokenizer, resulting in different tokenizations of the same text, and (2) code snippets, shell scripts, and output logs are summarized during issue report pre-processing, and these summaries can differ in length depending on the LLM used to generate them. 

\subsubsection{Findings} 

\begin{figure*}
        \centering
    \includegraphics[width=1.0\linewidth]{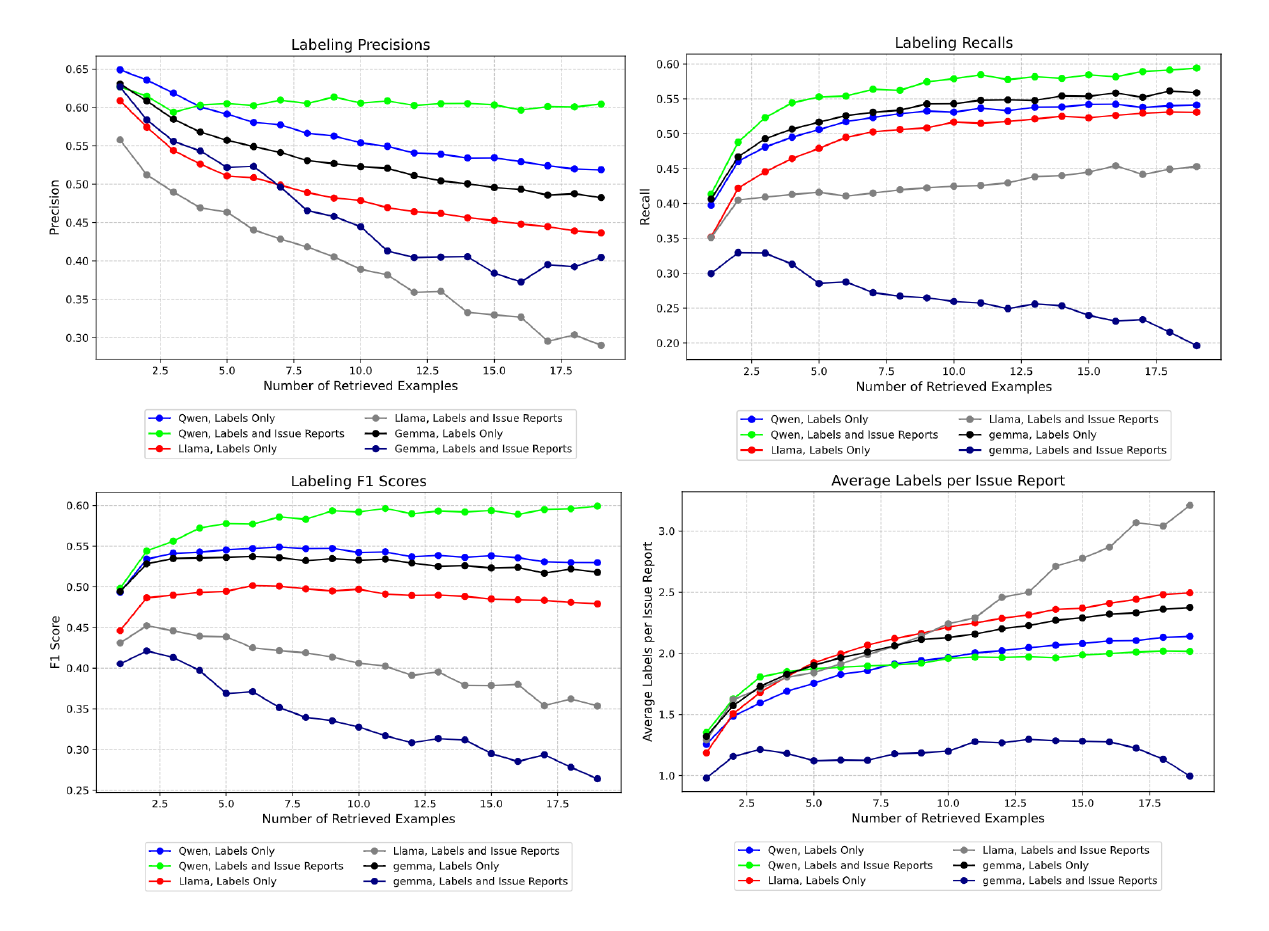}
    \caption{Alignment of original labels assigned to the 3,210 issue reports in the test set with those assigned by the LLMs to these issue reports using our two RAG-based prompts using various numbers of retrieved examples in terms of precision, recall, and F1 score, along with the average number of labels assigned to each issue report.
    }
    \label{fig:RAG_results_original_labels}
\end{figure*}

\begin{table}
\footnotesize
\centering
\caption{Alignment of original labels assigned to the 3,290 issue reports in the test set and labels assigned by our three label assigner LLMs to these issue reports using each of our three label assignment prompts, including each RAG-based prompt with the k-value with highest alignment to the original labels in terms of F1 score. Note that the average labels per issue report of the original labels is 2.05.}
\label{tab:assigned_labels_comparison_metrics}
\begin{tblr}{
  width = \linewidth,
  colspec = {Q[87]Q[163]Q[63]Q[52]Q[62]Q[152]Q[179]Q[177]},
  row{1} = {c,font=\bfseries},
  cell{2}{1} = {r=3}{font=\bfseries},
  cell{2}{3} = {c},
  cell{2}{4} = {c},
  cell{2}{5} = {c},
  cell{2}{6} = {c},
  cell{2}{7} = {c},
  cell{2}{8} = {c},
  cell{3}{3} = {c},
  cell{3}{4} = {c},
  cell{3}{5} = {c},
  cell{3}{6} = {c},
  cell{3}{7} = {c},
  cell{3}{8} = {c},
  cell{4}{3} = {c},
  cell{4}{4} = {c},
  cell{4}{5} = {c},
  cell{4}{6} = {c},
  cell{4}{7} = {c},
  cell{4}{8} = {c},
  cell{5}{1} = {r=3}{font=\bfseries},
  cell{5}{3} = {c},
  cell{5}{4} = {c},
  cell{5}{5} = {c},
  cell{5}{6} = {c,font=\bfseries},
  cell{5}{7} = {c},
  cell{5}{8} = {c},
  cell{6}{3} = {c},
  cell{6}{4} = {c},
  cell{6}{5} = {c},
  cell{6}{6} = {c},
  cell{6}{7} = {c},
  cell{6}{8} = {c},
  cell{7}{3} = {c},
  cell{7}{4} = {c},
  cell{7}{5} = {c},
  cell{7}{6} = {c},
  cell{7}{7} = {c},
  cell{7}{8} = {c},
  cell{8}{1} = {r=3}{font=\bfseries},
  cell{8}{3} = {c},
  cell{8}{4} = {c},
  cell{8}{5} = {c},
  cell{8}{6} = {c},
  cell{8}{7} = {c},
  cell{8}{8} = {c},
  cell{9}{3} = {c},
  cell{9}{4} = {c},
  cell{9}{5} = {c},
  cell{9}{6} = {c},
  cell{9}{7} = {c,font=\bfseries},
  cell{9}{8} = {c,font=\bfseries},
  cell{10}{3} = {c,font=\bfseries},
  cell{10}{4} = {c,font=\bfseries},
  cell{10}{5} = {c,font=\bfseries},
  cell{10}{6} = {c},
  cell{10}{7} = {c},
  cell{10}{8} = {c},
  hlines,
}
Label Assigner & Prompt                                & Precision & Recall & F1 Score & Average Labels per Issue Report & Average Runtime per Issue Report (s) & Average Number of Tokens per Prompt \\
Gemma          & Full Label List                       & 0.31      & 0.40   & 0.35     & 2.60                            & 0.72                                 & 1,903.08                            \\
               & RAG Labels only, k = 6                & 0.55      & 0.53   & 0.54     & 1.96                            & 0.49                                 & 442.80                              \\
               & RAG Labels and issue reports, k = 2   & 0.58      & 0.33   & 0.42     & 1.16                            & 12.07                                & 1133.15                             \\
Llama          & Full Label List                       & 0.28      & 0.47   & 0.35     & 3.40                            & 0.84                                 & 1,666.16                            \\
               & RAG Labels only, k = 6                & 0.51      & 0.49   & 0.50     & 1.91                            & 0.46                                 & 403.22                              \\
               & RAG Labels and issue reports, k = 2   & 0.51      & 0.40   & 0.45     & 1.62                            & 7.69                                 & 993.53                              \\
Qwen           & Full Label List                       & 0.46      & 0.41   & 0.43     & 1.84                            & 0.39                                 & 1,682.69                            \\
               & RAG Labels only, k = 7                & 0.58      & 0.52   & 0.55     & 1.86                            & 0.23                                 & 393.40                              \\
               & RAG labeled and issue reports, k = 19 & 0.61      & 0.59   & 0.60     & 2.02                            & 0.85                                 & 5,966.26                            
\end{tblr}
\end{table}

% \textbf{Labels assigned by LLMs to issue reports are moderately aligned with the original labels assigned to these issue reports.} As shown in Figure \ref{fig:RAG_results_original_labels} and Table \ref{tab:assigned_labels_comparison_metrics}, the LLM-based labeling configuration with the greatest alignment to the original labels is Qwen using the \textit{Labels only} prompt at $k=19$, achieving the highest precision, recall, and F1 scores. However, the values of these scores are only 0.61, 0.59, and 0.60 respectively. This indicates that the labels assigned by LLMs to issue reports often differ from the original labels assigned to these issue reports.

% \begin{figure*}
%         \centering
%     \includegraphics[width=1.0\linewidth]{figs/example_issue_reports.drawio.pdf}
%     \caption{Two issue reports in the test set of our dataset.
%     }
%     \label{fig:example_issue_reports}
% \end{figure*}

\begin{figure*}
        \centering
    \includegraphics[width=0.4\linewidth]{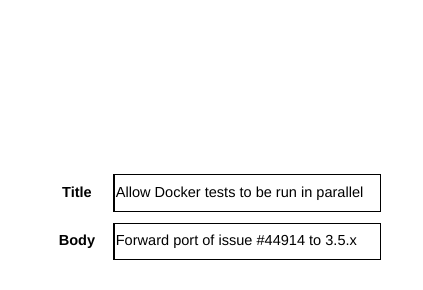}
    \caption{Title and body of issue report \#44915 in the repository \textit{spring-projects/spring-boot}.
    }
    \label{fig:example_issue_reports}
\end{figure*}

\begin{table}
\centering
\footnotesize
\caption{Comparison of original labels and the labels assigned by our three label assigner LLMs using each of our three label assignment prompts to issue report \#44915 in the repository \textit{spring-projects/spring-boot}.}
\label{tab:assigned_labels_comparison}
\begin{tblr}{
  width = \linewidth,
  colspec = {Q[171]Q[385]Q[387]},
  row{1} = {c,font=\bfseries},
  cell{2}{1} = {font=\bfseries},
  cell{2}{3} = {font=\itshape},
  cell{3}{1} = {r=3}{font=\bfseries},
  cell{3}{3} = {font=\itshape},
  cell{4}{3} = {font=\itshape},
  cell{5}{3} = {font=\itshape},
  cell{6}{1} = {r=3}{font=\bfseries},
  cell{6}{3} = {font=\itshape},
  cell{7}{3} = {font=\itshape},
  cell{8}{3} = {font=\itshape},
  cell{9}{1} = {r=3}{font=\bfseries},
  cell{9}{3} = {font=\itshape},
  cell{10}{3} = {font=\itshape},
  cell{11}{3} = {font=\itshape},
  hlines,
}
Label Assigner & Prompt                                & Assigned Labels                                        \\
Developers     & -                                     & \textit{\textit{type: task}}                           \\
Gemma          & Full Label List                       & \textit{\textit{type: enhancement, theme: containers}} \\
               & RAG Labels only, k = 6                & \textit{\textit{type: task}}                           \\
               & RAG Labels and issue reports, k = 2   & \textit{\textit{type: task}}                           \\
Llama          & Full Label List                       & \textit{\textit{type: enhancement, theme: containers}} \\
               & RAG Labels only, k = 6                & \textit{\textit{type: task}}                           \\
               & RAG Labels and issue reports, k = 2   & \textit{\textit{type: task}}                           \\
Qwen           & Full Label List                       & \textit{\textit{theme: containers, type: enhancement}} \\
               & RAG Labels only, k = 7                & \textit{\textit{type: task}}                           \\
               & RAG labeled and issue reports, k = 19 & \textit{\textit{type: task, theme: containers}}        
\end{tblr}
\end{table}

\textbf{The labels assigned by LLMs provide a more technically detailed description of the issue report than the original labels.} Figure \ref{fig:RAG_results_original_labels} shows the alignment of original labels assigned to the 3,210 issue reports in the test set with those assigned by the LLMs to these issue reports using our two RAG-based prompts across all tested numbers of retrieved examples (i.e., $k = 1$ to $k = 19$) in terms of precision, recall, and F1 score, along with the average number of labels assigned to each issue report. As shown in Figure \ref{fig:RAG_results_original_labels}, we see that the LLM-based labeling configuration with the greatest alignment to the original labels is Qwen using the \textit{Labels only} prompt at $k = 19$, achieving the highest precision, recall, and F1 scores of 0.61, 0.59, and 0.60 respectively. Table \ref{tab:assigned_labels_comparison_metrics} shows the alignment of original labels assigned to the 3,290 issue reports in the test set and labels assigned by our three label assigner LLMs to these issue reports using each of our three label assignment prompts, including each RAG-based prompt with the k-value with highest alignment to the original labels in terms of F1 score. From Table \ref{tab:assigned_labels_comparison_metrics}, we see that the LLM-based labeling configuration with the greatest alignment to the original labels is still Qwen using the Labels only prompt at $k=19$. This indicates that the labels assigned by LLMs to issue reports often differ from the original labels assigned to these issue reports. We then manually analyze the differences in assigned labels to determine whether the original labels or the labels assigned by our LLM-based labeling configurations provide a more technically detailed description of the issue report. Specifically, the leading author examined the labelings of 67 randomly sampled issue reports, a statistically representative subset of a 90\% confidence level and 10\% margin of error \cite{hazra2017using} of the 3,143 issue reports in our test set with original non-status labels (i.e., a status label refers to the the status of the issue’s resolution, e.g., \textit{help-wanted}, \textit{wontfix}, \textit{needs reproduction}). Of the 67 reviewed issue reports, we find that 24 (36\%) contain labels assigned by at least one LLM-based labeling configuration that provide a more technically detailed description of the concern reported in the issue report than the original labels. An example issue report where labels assigned by our LLM-based labeling configurations provide a more technically detailed description of the concern reported than the original labels is issue report \#44915 in the repository \textit{spring-projects/spring-boot}, whose title and body are shown in Figure \ref{fig:example_issue_reports}. As shown in Table \ref{tab:assigned_labels_comparison}, the only original label assigned to issue report \#44915 in \textit{spring-projects/spring-boot} is \textit{type:task}. In contrast, the labels assigned to issue report \#44915 in \textit{spring-projects/spring-boot} by the three LLM-based labeling configurations using the \textit{Full Label List} prompt are \textit{type: enhancement} and \textit{theme: containers}, both of which are recorded as false positives because these labels were not originally assigned to the issue report. However, as shown in Figure \ref{fig:example_issue_reports}, issue report \#44915 in \textit{spring-projects/spring-boot} is a request to allow Docker tests to be run in parallel. As a result, \textit{type: enhancement} and \textit{theme: containers} provide a more specific and accurate description of the issue report than \textit{type:task}. Specifically, \textit{type:task} only indicates that the issue involves work to be completed, whereas \textit{type: enhancement} conveys that the issue proposes an improvement to software functionality and \textit{theme: containers} identifies that the issue report relates to containerization. Since allowing Docker tests to be run in parallel constitutes an improvement to software functionality and Docker-related functionality falls under the theme of containers, the labels \textit{type: enhancement} and \textit{theme: containers}, as assigned by three of our LLM-based labeling configurations, provide a more technically detailed description of the concern reported in issue report \#44915 in \textit{spring-projects/spring-boot} than the original label \textit{type:task}. This finding demonstrates that evaluating automated labeling approaches solely against the original labels may underestimate their performance because the original labels may not represent the most detailed characterization of the concerns reported in issue reports. 

In our replication package\footnote{\href{https://github.com/24rrvk/LLMIssueLabeling/blob/main/Results\_and\_Prompts/LLM\_assigned\_original\_labels\_comparison/random\_sample\_w\_llm\_assigned\_original\_labels.csv}{https://github.com/24rrvk/LLMIssueLabeling/blob/main/Results\_and\_Prompts/LLM\_assigned\_original\_labels\_comparison/random\_sample\_w\_llm\_\\assigned\_original\_labels.csv}}, we make the following information available: (1) the titles and bodies of issue reports in this representative subset along with their original labels, (2) labels assigned by our LLM-based labeling configurations, and (3) whether the leading author determined if the labels assigned by one LLM-based labeling configuration more accurately described the concern reported in the issue report than the original labels.

\begin{figure*}
        \centering
    \includegraphics[width=1.0\linewidth]{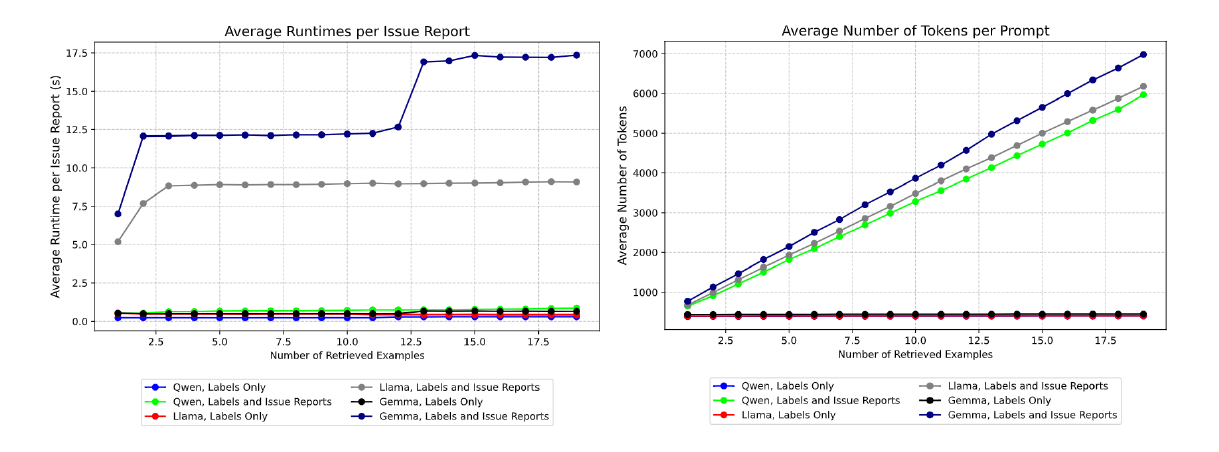}
    \caption{Average runtime per issue report and number of tokens per prompt using our two RAG-based prompts across various numbers of retrieved examples.
    }
    \label{fig:cost_runtime_analysis_MS1}
\end{figure*}

\textbf{\textit{Labels only} is the most efficient prompt in terms of runtime and prompt length.} As shown in Figure \ref{fig:cost_runtime_analysis_MS1}, the average runtime per issue report and average number of tokens per prompt for the \textit{Labels only} prompt remains relatively stable as the number of retrieved examples increases. This is because increasing the number of retrieved examples only increases the number of labels added to the \{label\_list\} placeholder in the prompt. In contrast, the \textit{Labels and issue reports} prompt also includes the content of the retrieved issue reports, so increasing the number of retrieved examples also increases the number of tokens in the prompt, which in turn increases the runtime. As shown in Table \ref{tab:assigned_labels_comparison_metrics}, the \textit{Labels only} prompt also uses fewer tokens than the \textit{Full Label List} prompt because it includes only the labels assigned to the retrieved examples, whereas the \textit{Full Label List} prompt includes the repository's entire label list. As a result, the \textit{Labels only} prompt is also able to maintain a shorter runtime than the \textit{Full Label List} prompt despite the fact that it needs to retrieve examples from the RAG database whereas the \textit{Full Label List} prompt does not.

\begin{Summary}[ ]{Summary of MS1}

The original labels assigned to issue reports often differ from those assigned by LLM-based issue report labeling systems when assigning from the same set of labels. However, the labels assigned by LLMs can provide a more accurate description of the issue report than the original labels assigned to these issue reports. As a result, treating the original labels assigned to issue reports as ground truth is problematic when evaluating the performance of LLM-based issue report labeling systems. 

\end{Summary}

\subsection{MS2: How do LLMs assign labels to issue reports in the absence of a pre-defined list?}

\subsubsection{Motivation} 
The motivation behind this study is to explore how LLMs behave when assigning labels without pre-defined constraints, thereby revealing their inherent labeling tendencies. This investigation is crucial because previous studies evaluate LLMs within overly generic labeling spaces (i.e., they restrict LLMs to assigning one of the broad labels \textit{bug}, \textit{feature}, \textit{question}, or \textit{documentation} to issue reports \cite{aracena2024applying, colavito2024leveraging}). By examining their natural labeling behavior, we aim to assess the extent to which LLMs can autonomously identify and categorize issue reports. To the best of our knowledge, no prior work has explored this aspect and the resulting insights directly inform the design of \textit{LabelMate}.

\subsubsection{Approach}

For each of the 13,210 issue reports in the training set, we prompt each of our three \textit{label assigner LLMs} to generate labels that categorize the issue's type and describe its domain. The template for this prompt is shown in Figure \ref{fig:label_gen_prompt}. Specifically, we design the prompt as follows:

% We prompt LLMs to generate labels for historical issue reports to serve as candidates for inclusion in our label list. 

\begin{figure*}
        \centering
    \includegraphics[width=1.0\linewidth]{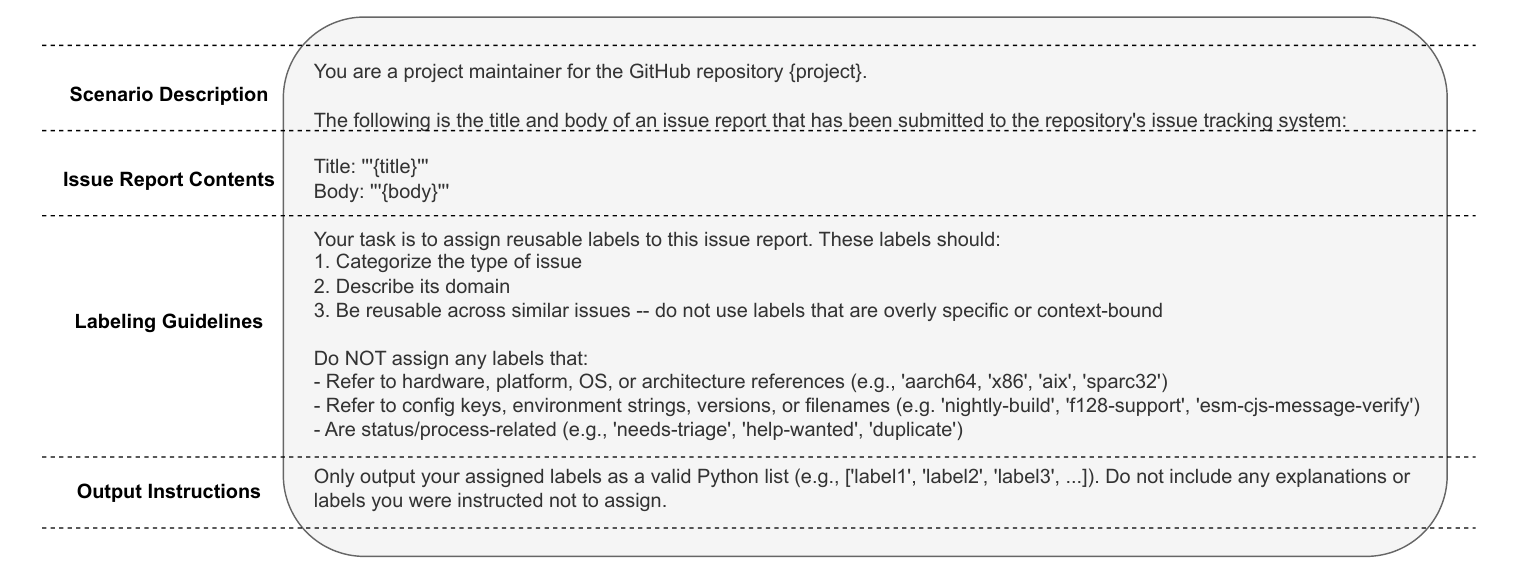}
    \caption{Label generation prompt.}
    \label{fig:label_gen_prompt}
\end{figure*}

% Below is a description and justification of each of the four sections of the prompt template:

\begin{itemize}
    \item \textbf{Scenario Description:} We position the LLM as a project maintainer for the GitHub repository to which the issue report has been submitted. For example, if the issue report belongs to the GitHub repository \textit{microsoft/vscode}\footnote{\href{https://github.com/microsoft/vscode}{https://github.com/microsoft/vscode}}, we fill the ``\{project\}'' placeholder in the prompt template with \textit{microsoft/vscode}.
    \item \textbf{Issue Report Contents:} We provide the LLM with both the \textit{title} and \textit{body} of each issue report, following prior work on automated issue report labeling with LLMs~\cite{aracena2024applying, colavito2024leveraging}. The title typically offers a concise summary of the problem or requested feature, whereas the body elaborates on technical context, reproduction steps, observed behavior, or proposed solutions.
    \item \textbf{Labeling Guidelines:} We include explicit labeling guidelines in the prompt to steer the LLM toward generating relevant, reusable, and semantically meaningful labels (e.g., those describing issue type or domain). Without such guidance, LLMs frequently output overly specific or irrelevant terms (e.g., hardware architectures, configuration keys). Negative examples of these undesirable labels are provided to prevent them from being outputted (e.g., \textit{aarch64}, \textit{f128-support}), whereas positive examples are omitted to avoid constraining the model’s creativity and ensure broad thematic coverage across issues.
    \item \textbf{Output Instructions:} We instruct the LLM to output the labels it assigns as a Python list and to not include any explanations to ensure ease of processing of the outputted labels.
\end{itemize}

To standardize the raw labels produced by the LLM, we apply \textit{output normalization steps} to remove surface-level variations and duplicates. Specifically, we convert all text to lowercase and replace delimiters (e.g., underscores, hyphens) with spaces. For example, the labels \textit{Feature request}, \textit{feature\_request}, and \textit{Feature-Request} are normalized to \textit{feature request}. This step ensures consistent representation of labels while preserving their semantic content.

\begin{table}
\footnotesize
\centering
\caption{Summary statistics of labels generated by the three label assigner LLMs from the 13,210 issue reports in the training set.}
\label{tab:label_gen_stats}
\begin{tblr}{
  width = \linewidth,
  colspec = {Q[169]Q[192]Q[242]Q[335]},
  column{even} = {c},
  column{1} = {font=\bfseries},
  column{3} = {c},
  hlines,
}
LLM           & {\textbf{Number of~Unique }\\\textbf{Labels~Generated}} & {\textbf{Number of~Generated }\\\textbf{Labels~per Issue Report}} & {\textbf{Percentage of Unique~Labels }\\\textbf{Generated~from One Issue~Report}} \\
Gemma         & 5,925                                                   & 2.72                                                              & 59.78\%                                                                           \\
Llama         & 8,020                                                   & 4.94                                                              & 61.77\%                                                                           \\
Qwen          & 7,552                                                   & 2.97                                                              & 63.90\%                                                                           \\
LLMs Combined & 14,960                                                  & 8.14                                                                & 63.05\%                                                                                
\end{tblr}
\end{table}

\subsubsection{Findings} 

%\indent\hspace*{\parindent}\textit{\textbf{LLMs generate an excessive number of labels.}} 

\textbf{LLMs generate an excessive number of labels.} As shown in Table~\ref{tab:label_gen_stats}, the models produce between 5,925 and 8,020 unique labels from 13,210 issue reports. Such large, fragmented label spaces are impractical for real-world use: they complicate issue filtering, restrict label reuse, and undermine the organizational purpose of labeling systems. In practice, an inflated label set would make navigation and retrieval of related issues inefficient, thereby reducing the utility of labeling in software maintenance and project management.

\textbf{LLMs generate overly-specific, non-generalizable labels.} As shown in Table \ref{tab:label_gen_stats}, across all models, at least 59\% of the unique labels are generated from only a single issue report. This behavior indicates that many generated labels are too narrow to serve a grouping function, which contradicts the primary goal of labeling of facilitating the categorization and retrieval of related issues. Such hyper-specific labels may capture incidental details rather than generalizable issue types.

\textbf{LLMs generate different labels with similar meaning, or synonymous labels.} Another major limitation we observe with this approach is the generation of \textbf{synonymous labels}, i.e., different labels that refer to the same underlying concept, for different issue reports. While surface-level variations can be handled by our output normalization steps (e.g., when applying these steps to the labels \textit{User-Interface} and \textit{user\_interface}, they both become \textit{user interface}), labels with different textual representation but almost identical semantic meaning cannot.  Examples of such labels generated in this study are \textit{ui} and \textit{user interface} and \textit{enhancement} and \textit{improvement}. Including this redundancy in label sets could lead to the fragmentation of conceptually related issues across multiple synonymous labels, and therefore lead to an undermining of the consistency of the resulting categorizations. The extent of this synonymy is systematically analyzed in Section \ref{sec:rq1}, where we cluster and refine these labels to construct a coherent label taxonomy.

\textbf{Even when using the same prompt, LLMs often generate different labels for the same issue report.} When combining the labels generated by the three label assigner LLMs for each issue report, the average number of labels generated per issue report is 8.14. Additionally, an average of only 2.49 labels per issue report were generated by more than one model, meaning that an average of 5.65 labels per issue report were only generated by a single model. This suggests that the LLMs frequently generate different labels for the same issue report, even when given the same prompt.

\begin{Summary}[ ]{Summary of MS2}
LLMs exhibit three key limitations when assigning labels without a pre-defined list: they tend to (1) generate excessively large and fragmented label spaces, (2) produce many labels that apply to only a single issue report, and (3) create redundant or synonymous terms.
\end{Summary}

\section{Results}
\label{sec:rqs}

This section describes the motivation, approaches, and findings of each of our four research questions (RQs).

% \subsection{RQ1: How can the labels generated by LLMs from issue reports be consolidated into a coherent label list?}
% \label{sec:rq2}

\subsection{RQ1: How can we derive a coherent label list for a set of issue reports?}
\label{sec:rq1}

\subsubsection{Motivation} 

Findings from our second motivational study reveal that allowing LLMs to assign labels without a pre-defined list leads to an incoherent labeling space characterized by redundancy, over-specificity, and excessive diversity. Such fragmentation undermines the main purpose of labeling—enabling consistent organization and retrieval of issue reports.

To address these challenges, we propose constraining LLMs to a pre-defined label list. A fixed list promotes consistency, reusability, and interpretability of labels across issues. However, defining such a list manually is both labour-intensive and context-dependent, which limits scalability and adoption in practice~\cite{fan2017road}. Instead, we leverage the labels generated in the motivational study as a foundation and develop a systematic consolidation process that refines these labels into a coherent, representative list.

\subsubsection{Approach} 
\label{sec:deriving_label_list_approach}

To produce a coherent and non-redundant label list, we consolidate the 14,960 unique labels generated in MS2 through the following four steps:

\textbf{Step 1: Process Labels.} To ensure that the labels in our generated list are reusable across issue reports, we exclude labels generated by only a single model for a single issue report. This filtering step removes 8,788 labels, leaving 6,172 candidates for consolidation. Given that in our second motivational study (i.e., MS2) we generate labels from 13,210 issue reports, labels generated by only a single model for a single issue report do not represent reusable or generalizable categories. Instead, they typically correspond to overly specific descriptions or artifacts, e.g., \textit{nextjsdev}, \textit{go111module}, and \textit{platform plan9}, which reflect narrow contextual details rather than recurring issue types. Retaining all low frequency and overly-specific labels would substantially increase the sparsity and noise in the label space, leading to a highly fragmented taxonomy that hinders downstream consolidation and reduces interpretability of the generated taxonomy. % While this may exclude labels that describe specific details of the issue, the objective of this research question is to derive a coherent list of issue report categories that capture recurring issue types, rather than exhaustively preserving every characteristic of all issues. 
Consequently, we only retain labels that were generated either from more than one issue report or by more than one LLM in MS2 for categorical stability and reusability. 

\textbf{Step 2: Semantic Grouping of Labels.} The first step in this process is to compute embeddings of each candidate label in our set of 6,172 using our text embedding model \textit{all-mpnet-base-v2} to quantitatively capture their semantic meaning \cite{chersoni2021decoding}. We then group semantically similar labels by clustering their embeddings using \textit{Agglomerative Clustering} \cite{mullner2011modern}, a hierarchical algorithm that iteratively merges the closest clusters until a termination condition is reached. This method has proven effective in grouping semantically related text while preserving conceptual distinctions \cite{petukhova2025text, saha2023influence, sajeva2024clustering}.
Three parameters control the algorithm: the distance metric, linkage criterion, and termination condition. We adopt cosine distance (i.e., the angular separation between two embedding vectors) as the distance metric and average linkage (i.e., the mean distance between all pairs of points in the two clusters) as the linkage criterion. Clustering stops when no pair of clusters has a distance exceeding the threshold $\tau$.  We conduct a sensitivity analysis on the distance threshold parameter $\tau \in {0.2, 0.3, 0.4, 0.5, 0.6, 0.7, 0.8}$ to determine the configuration that best balances conceptual granularity and semantic cohesion. Summary statistics of the resulting clusters are shown in Table~\ref{tab:mpnet_clustering_stats}; full clusters are available in the replication package\footnote{\href{https://github.com/24rrvk/LLMIssueLabeling/tree/main/Results_and_Prompts/clusters/all-mpnet-base-v2_metric\%3Dcos_link\%3Davg}{https://github.com/24rrvk/LLMIssueLabeling/tree/main/Results\_and\_Prompts /clusters/all-mpnet-base-v2\_metric\%3Dcos\_link\%3Davg}}.

\begin{table*}
\centering
\footnotesize
\caption{Summary statistics of clustering the 6,172 labels at various distance thresholds.}
\begin{tblr}{
  width = \linewidth,
  colspec = {Q[119]Q[123]Q[121]Q[162]Q[196]Q[213]},
  cells = {c},
  hlines,
}
{\textbf{Distance}\\\textbf{Threshold}} & {\textbf{Number of}\\\textbf{Clusters}} & {\textbf{Labels per}\\\textbf{Cluster}} & {\textbf{Median Labels}\\\textbf{per Cluster}} & {\textbf{Number of Labels}\\\textbf{in Largest Cluster}} & {\textbf{Number of Clusters}\\\textbf{with one Label}} \\
0.2                                     & 165                                     & 37.41                                   & 23                                             & 347                                                      & 0                                                      \\
0.3                                     & 529                                     & 11.67                                   & 8                                              & 118                                                      & 22                                                     \\
0.4                                     & 1,036                                   & 5.96                                    & 4                                              & 98                                                       & 164                                                    \\
0.5                                     & 1,678                                   & 3.68                                    & 2                                              & 56                                                       & 507                                                    \\
0.6                                     & 2,691                                   & 2.29                                    & 2                                              & 43                                                       & 1,296                                                  \\
0.7                                     & 3,953                                   & 1.56                                    & 1                                              & 21                                                       & 2,645                                                  \\
0.8                                     & 4,976                                   & 1.24                                    & 1                                              & 12                                                       & 4,042                                                  
\end{tblr}
\label{tab:mpnet_clustering_stats}
\end{table*}

When observing these clusters, we find that at thresholds below 0.3, conceptually distinct labels appear in the same clusters (e.g., \textit{regression}, \textit{quantization}, \textit{image processing}), while thresholds above 0.3 lead to unnecessary fragmentation (e.g., separating \textit{documentation} and \textit{docs}, or \textit{segmentation fault} and \textit{segfault}). Hence, a threshold of 0.3 provides the best trade-off between cohesion and differentiation. Therefore, we conclude that the optimal set of clusters is yielded when the distance threshold is set to 0.3. 

\textbf{Step 3: Representative Label Selection.} For of the 529 clusters yielded from the optimal clustering configuration, we select a single representative label to capture its underlying concept. Specifically, the label most frequently generated during candidate generation is chosen, as frequency reflects its prominence in developer discourse. This ensures that only one label per concept is retained, thereby reducing redundancy and preventing synonymous terms from inflating the list. For example, the labels \textit{enhancement}, \textit{improvement}, \textit{suggestion}, and \textit{proposal} were grouped together in the optimal clustering configuration. Since \textit{enhancement} was the label most frequently generated, it is selected as the cluster’s representative.

\textbf{Step 4: Representative Label Evaluation and Filtering.} Upon inspection of the 529 representative labels, we observe that many are \textbf{names of tools or infrastructure components} (e.g., programming languages such as \textit{java} or \textit{css}; operating systems such as \textit{windows} or \textit{android}; compilers such as \textit{rustc}; and databases such as \textit{redis} or \textit{tidb}). % Remember the text pre-processing steps!!! These are all lowercased!!!!!!
Such labels are commonly used in the label taxonomies of real-world repositories. For example, the GitHub repository \textit{angular/angular}\footnote{\url{https://github.com/angular/angular}} uses labels such as \textit{browser: chrome}, \textit{browser: firefox}, \textit{browser: safari} to denote the browser in which the issue occurs.  However, labels referring to the names of specific tools or infrastructure components are tightly coupled to the scope and technology stack of the individual project. The objective of our generated taxonomy is to provide a reusable set of labels that can be broadly adopted across collaborative software repositories of a variety of domains. Consequently, retaining labels that are names of tool or infrastructure components would reduce the generalizability of our taxonomy across software domains. Therefore, we instead retain the more general forms of labels that are names of tools or infrastructure components. For example, instead of retaining labels referring to a project's specific compiler, e.g., \textit{rustc}, we retain the generalizable label \textit{compiler}. Similarly, rather than retaining labels referring to the names of specific databases, e.g., \textit{redis} and \textit{tidb}, we retain the generalizable label \textit{database}. While the generalizable labels do not communicate the exact technologies involved in the issue, they preserve the functional and conceptual nature of the issue while remaining applicable across a broader range of collaborative software repositories.  \\

After these consolidation steps, we obtain a coherent label list excluding redundant and non-informative terms. 

\subsubsection{Findings} 
\label{sec:rq1_findings}

\textbf{There is a high degree of synonyms in the labels generated by LLMs in the absence of a pre-defined list.} Given that we find the optimal set of clusters is yielded when the distance threshold is set to 0.3, Table \ref{tab:mpnet_clustering_stats} shows that this configuration results in average and median labels per cluster of 11.67 and 8, respectively, with only 22 of the 529 clusters containing a single label. These results indicate that the vast majority of labels share substantial semantic overlap as multiple labels can be grouped together under a single conceptual cluster. This high clustering density demonstrates that LLMs tend to produce numerous variations of the same underlying concept when not constrained by a pre-defined label list. This reinforces the necessity of providing LLMs with a fixed, coherent list of labels to ensure consistency and reduce redundancy in issue report categorization.

\textbf{Nearly half of representative labels lack descriptive value.} Two authors independently examined each of the 529 representative labels and classified them as either descriptive (conceptual category) or non-descriptive (e.g., names of tools, frameworks, or programming languages). Agreement between authors was high with a Cohen's Kappa value \cite{cohen1960coefficient} of 0.8766, an \textit{Almost Perfect} strength of agreement according to Landis and Koch \cite{landis1977measurement}. For disagreements, the authors discussed and reached a consensus on inclusion. Ultimately, 254 labels are excluded as non-descriptive, resulting in a final coherent list of 275 labels. All classifications and adjudications are publicly available in the replication in the replication package\footnote{\href{https://github.com/24rrvk/LLMIssueLabeling/blob/main/Results_and_Prompts/label_list}{https://github.com/24rrvk/LLMIssueLabeling/blob/main/Results\_and\_Prompts/label\_list}}.

\begin{Summary}[ ]{Summary of RQ1}
By consolidating 14,960 noisy and redundant labels generated by three LLMs through semantic clustering and representative selection, we derive a coherent taxonomy of 275 descriptive labels.
\end{Summary}

\subsection{RQ2: How do LLMs assign labels to issue reports from our coherent label list?}
\label{sec:rq3}

\subsubsection{Motivation} 

Building on the findings of RQ1, where we derived a coherent list of 275 descriptive labels, this research question investigates how LLMs behave when constrained to assigning labels exclusively from this list. While prior studies have evaluated LLMs in narrow labeling settings, typically limited to broad categories such as \textit{bug}, \textit{feature}, \textit{question}, or \textit{documentation}~\cite{aracena2024applying, colavito2024leveraging}, real-world software projects exhibit far greater thematic diversity. By analyzing LLM behaviour under this more realistic constraint, we can evaluate how effectively they can leverage a structured label space to produce meaningful assignments. This provides insights that go beyond prior studies focusing on overly coarse label categories \cite{aracena2024applying, colavito2024leveraging}. 

\subsubsection{Approach} 
\label{sec:rq3_approach}

We evaluate the labeling performance of each of the three label assigner LLMs, i.e., \textit{Gemma}, \textit{Llama}, and \textit{Qwen}, on the 13,210 issue reports in the training set. Each model is prompted to assign labels from the list of 275 labels generated in RQ1 using prompt shown in Figure \ref{fig:full_label_list_and_RAG_labels_only_prompt}, where the list of 275 labels generated in RQ1 is inserted in the \{label\_list\} placeholder. For comparison, we reproduce three baseline settings where models can assign one label to an issue report from the following label lists:

\begin{enumerate}
    \item \textbf{Colavito et al. Label List:} A standard list of the four coarse-grained labels \textit{bug}, \textit{feature}, \textit{question}, and \textit{documentation} used in \cite{colavito2024leveraging}.
    \item \textbf{Catolino et al. Label List:} A manually derived list of the 9 labels \textit{Configuration issue}, \textit{Database-related issue}, \textit{GUI-related issue}, \textit{Network issue}, \textit{Performance issue}, \textit{Permission/Deprecation issue}, \textit{Program Anomaly issue}, \textit{Security issue}, and \textit{Test Code-related issue}.
    \item \textbf{Assi et al. Label List:} A list of 15 labels derived through Embedded Topic Modeling. These labels are \textit{Platform compatibility}, \textit{Testing}, \textit{User experience}, \textit{File management}, \textit{Build and deployment}, \textit{API related issues}, \textit{Security}, \textit{Release and Update}, \textit{Performance}, \textit{Database}, \textit{Parallel event processing}, \textit{General program related anomaly}, \textit{GUI}, \textit{Server issues}, and \textit{Interprocess communication (IPC)}.
\end{enumerate}

The prompt used for these baselines is a modified version of the prompt shown in Figure \ref{fig:full_label_list_and_RAG_labels_only_prompt} in which ``assign the most appropriate label(s) for this issue report in the form of a Python list (e.g., ['label1', 'label2', 'label3', ...]'' is replaced with ``assign the most appropriate label for this issue report''. This modification restricts the LLM to assigning a single label from the provided lists to maintain consistency with the classifiers used in these prior works, which were also limited to single-label assignment.

We also compare with the original labels assigned to the issue reports by developers. To ensure a fair comparison, we remove labels that do not directly refer to the issue itself. These include labels that refer to the \textit{status of the issue's resolution} (e.g., \textit{help-wanted}, \textit{wontfix}, \textit{needs reproduction}), labels that \textit{compare issues to other issues} (e.g., \textit{good first issue}, \textit{duplicate}, \textit{high priority}), and \textit{version identification labels} (e.g., \textit{Vuetify 2}, \textit{affects-7.6}). The classification of these labels as either directly referring to or not directly referring to the issue itself can be viewed in our replication package\footnote{\url{https://github.com/24rrvk/LLMIssueLabeling/tree/main/Results_and_Prompts/original_label_lists/projects_in_dataset}}. Lastly, we employ the following ensembling techniques to determine if combining label assignments from multiple LLMs improves labeling performance: 

\begin{itemize}
    \item \textbf{Unanimous voting among each pair of LLMs} considers labels that both LLMs in the pair assigned to the same issue report.
    \item \textbf{Unanimous voting among all three LLMs} considers labels assigned by all three LLMs to the same issue report.
    \item \textbf{Majority voting among all three LLMs} considers labels assigned by at least two LLMs to the same issue report.
\end{itemize}

\textbf{Evaluation Metrics.} In the first motivational study, we found that in 24 of 67 issue reports (36\%), labels assigned by at least one LLM-based labeling configuration more accurately describe the concern reported in the issue report than the original labels. As a result, we evaluate label assignments by measuring the semantic alignment between assigned labels and issue report content using cosine similarity, a standard metric for assessing semantic relatedness between text embeddings~\cite{turney2010frequency}. The rationale for using this metric is that embeddings encode semantic information such that semantically related texts are positioned closer together in the embedding space. As a result, if the embedding of an assigned label is more similar to the embedding of the corresponding issue report content, this indicates stronger semantic correspondence between the two. Consequently, a higher cosine similarity score suggests that the assigned label more accurately captures the underlying meaning and context of the issue report and thus provides a better semantic description of it.

In terms of how we specifically measure semantic alignment, we embed both the issue report (title and body) and its assigned labels using our text embedding model \textit{all-mpnet-base-v2}, and compute the cosine similarity between each embedded label and the embedded issue content. For example, the original developer-assigned labels for issue \#26653 of the project \textit{denoland/deno}\footnote{\url{https://github.com/denoland/deno/issues/26653}} are \textit{bug} and \textit{install}. These labels have cosine similarity values of 0.1192 and 0.1668 respectively with the embedded issue report content. The semantic alignment of the assigned labels with the issue report is then determined by taking the average similarity across all labels--in this case, $(0.1192+0.1668)/2 = 0.1430$. Note that if no labels are assigned to an issue report, the similarity is recorded as 0, reflecting the absence of labeling effort. These values are averaged for all issue reports to obtain a labeling configuration's \textit{Average Cosine Similarity}. 

We also report Cohen's Kappa values \cite{cohen1960coefficient} for each pair of LLMs across all label sets to quantify inter-model agreement. To calculate Cohen's Kappa for model assignments from our derived label list, where any number of labels could be assigned to a given issue report, each model's assignments are represented as a one-dimensional binary array \textit{x} of length $13,210\times275$. Since there are 13,210 issue reports and 275 possible labels, if the issue report \textit{i} is assigned the label \textit{j}, then $x_{275i+j}=1$; otherwise $x_{275i+j}=0$. Here, \textit{i} indexes the issue reports (from 0 to 13,209) and \textit{j} indexes the labels (from 0 to 274). The Cohen's Kappa value is then computed between the two resulting binary arrays.

\subsubsection{Findings} 

\begin{table}
\centering
\footnotesize
\caption{Average cosine similarity of original developer-assigned labels and LLM-assigned labels selected from either the Colavito et al., Catolino et al., or Assi et al. label lists, or our derived list of labels with the 13,210 issue reports in the training set.}
\label{tab:big_results_table}
\begin{tblr}{
  width = \linewidth,
  colspec = {Q[238]Q[127]Q[175]Q[162]Q[135]Q[98]},
  row{1} = {c},
  column{1} = {font=\bfseries},
  cell{1}{3} = {font=\bfseries},
  cell{2}{4} = {c},
  cell{2}{5} = {c},
  cell{2}{6} = {c},
  cell{3}{1} = {r=4}{},
  cell{3}{2} = {r=4}{},
  cell{3}{4} = {c},
  cell{3}{5} = {c},
  cell{3}{6} = {c},
  cell{4}{4} = {c},
  cell{4}{5} = {c},
  cell{4}{6} = {c},
  cell{5}{4} = {c},
  cell{5}{5} = {c},
  cell{5}{6} = {c},
  cell{6}{4} = {c},
  cell{6}{5} = {c},
  cell{6}{6} = {c},
  cell{7}{1} = {r=4}{},
  cell{7}{2} = {r=4}{},
  cell{7}{4} = {c},
  cell{7}{5} = {c},
  cell{7}{6} = {c},
  cell{8}{4} = {c},
  cell{8}{5} = {c},
  cell{8}{6} = {c},
  cell{9}{4} = {c},
  cell{9}{5} = {c},
  cell{9}{6} = {c},
  cell{10}{4} = {c},
  cell{10}{5} = {c,font=\bfseries},
  cell{10}{6} = {c},
  cell{11}{1} = {r=4}{},
  cell{11}{2} = {r=4}{},
  cell{11}{4} = {c},
  cell{11}{5} = {c},
  cell{11}{6} = {c},
  cell{12}{4} = {c},
  cell{12}{5} = {c},
  cell{12}{6} = {c},
  cell{13}{4} = {c},
  cell{13}{5} = {c},
  cell{13}{6} = {c},
  cell{14}{4} = {c,font=\bfseries},
  cell{14}{5} = {c},
  cell{14}{6} = {c},
  cell{15}{1} = {r=4}{},
  cell{15}{2} = {r=4}{},
  cell{15}{4} = {c},
  cell{15}{5} = {c},
  cell{15}{6} = {c},
  cell{16}{4} = {c},
  cell{16}{5} = {c},
  cell{16}{6} = {c},
  cell{17}{4} = {c},
  cell{17}{5} = {c},
  cell{17}{6} = {c},
  cell{18}{4} = {c},
  cell{18}{5} = {c},
  cell{18}{6} = {c},
  cell{19}{1} = {r=4}{},
  cell{19}{2} = {r=4}{},
  cell{19}{4} = {c},
  cell{19}{5} = {c},
  cell{19}{6} = {c},
  cell{20}{4} = {c},
  cell{20}{5} = {c},
  cell{20}{6} = {c},
  cell{21}{4} = {c},
  cell{21}{5} = {c},
  cell{21}{6} = {c},
  cell{22}{4} = {c},
  cell{22}{5} = {c},
  cell{22}{6} = {c},
  cell{23}{1} = {r=4}{},
  cell{23}{2} = {r=4}{},
  cell{23}{4} = {c},
  cell{23}{5} = {c},
  cell{23}{6} = {c},
  cell{24}{4} = {c},
  cell{24}{5} = {c},
  cell{24}{6} = {c},
  cell{25}{4} = {c},
  cell{25}{5} = {c},
  cell{25}{6} = {c},
  cell{26}{4} = {c},
  cell{26}{5} = {c},
  cell{26}{6} = {c},
  cell{27}{1} = {r=4}{},
  cell{27}{2} = {r=4}{},
  cell{27}{4} = {c},
  cell{27}{5} = {c},
  cell{27}{6} = {c},
  cell{28}{4} = {c},
  cell{28}{5} = {c},
  cell{28}{6} = {c},
  cell{29}{4} = {c},
  cell{29}{5} = {c},
  cell{29}{6} = {c},
  cell{30}{4} = {c},
  cell{30}{5} = {c},
  cell{30}{6} = {c},
  cell{31}{1} = {r=4}{},
  cell{31}{2} = {r=4}{},
  cell{31}{4} = {c},
  cell{31}{5} = {c},
  cell{31}{6} = {c},
  cell{32}{4} = {c},
  cell{32}{5} = {c},
  cell{32}{6} = {c},
  cell{33}{4} = {c},
  cell{33}{5} = {c},
  cell{33}{6} = {c},
  cell{34}{4} = {c},
  cell{34}{5} = {c},
  cell{34}{6} = {c},
  hlines,
}
Label Assigner       & {\textbf{Ensembling}\\\textbf{Technique}} & Label List          & {\textbf{Average Cosine }\\\textbf{Similarity}} & {\textbf{Labels per }\\\textbf{Issue Report}} & {\textbf{Cohen's}\\\textbf{ Kappa}} \\
Developer   & -                                         & Project Label List  & 0.171                                           & 1.89                                          & -                                   \\
Gemma                & -                                         & Colavito et al. Label List    & 0.135                                           & 0.99                                          & -                                   \\
                     &                                           & Catolino et al. Label List & 0.141                                           & 0.98                                          & -                                   \\
                     &                                           & Assi et al. Label List     & 0.108                                           & 0.99                                          & -                                   \\

                     &                                           & Derived Label List  & 0.162                                           & 3.32                                          & -                                   \\
Llama                & -                                         & Colavito et al. Label List     & 0.128                                           & 1.00                                          & -                                   \\
                     &                                           & Catolino et al. Label List & 0.143                                           & 1.00                                          & -                                   \\
                     &                                           & Assi et al. Label List     & 0.095                                           & 1.00                                          & -                                   \\

                     &                                           & Derived Label List  & 0.147                                           & 3.72                                          & -                                   \\
Qwen                 & -                                         & Colavito et al. Label List     & 0.136                                           & 1.00                                          & -                                   \\
                     &                                           & Catolino et al. Label List & 0.133                                           & 0.97                                          & -                                   \\
                     &                                           & Assi et al. Label List     & 0.133                                           & 1.00                                          & -                                   \\

                     &                                           & Derived Label List  & 0.178                                           & 2.31                                          & -                                   \\
Gemma + Llama        & Unanimous                                 & Colavito et al. Label List     & 0.123                                           & 0.83                                          & 0.67                                \\
                     &                                           & Catolino et al. Label List & 0.103                                           & 0.66                                          & 0.60                                \\
                     &                                           & Assi et al. Label List     & 0.075                                           & 0.65                                          & 0.61                                \\

                     &                                           & Derived Label List  & 0.173                                           & 2.30                                          & 0.65                                \\
Gemma + Qwen         & Unanimous                                 & Colavito et al. Label List     & 0.129                                           & 0.91                                          & 0.82                                \\
                     &                                           & Catolino et al. Label List & 0.107                                           & 0.69                                          & 0.60                                \\
                     &                                           & Assi et al. Label List     & 0.086                                           & 0.64                                          & 0.60                                \\

                     &                                           & Derived Label List  & 0.176                                           & 1.65                                          & 0.58                                \\
Llama + Qwen         & Unanimous                                 & Colavito et al. Label List     & 0.123                                           & 0.83                                          & 0.68                                \\
                     &                                           & Catolino et al. Label List & 0.091                                           & 0.56                                          & 0.48                                \\
                     &                                           & Assi et al. Label List     & 0.072                                           & 0.56                                          & 0.52                                \\

                     &                                           & Derived Label List  & 0.173                                           & 1.58                                          & 0.52                                \\
Gemma + Llama + Qwen & Majority                                  & Colavito et al. Label List     & 0.135                                           & 0.98                                          & -                                   \\
                     &                                           & Catolino et al. Label List & 0.138                                           & 0.92                                          & -                                   \\
                     &                                           & Assi et al. Label List     & 0.107                                           & 0.89                                          & -                                   \\

                     &                                           & Derived et al. Label List  & 0.174                                           & 2.76                                          & -                                   \\
Gemma + Llama + Qwen & Unanimous                                 & Colavito et al. Label List     & 0.121                                           & 0.80                                          & -                                   \\
                     &                                           & Catolino et al. Label List & 0.082                                           & 0.50                                          & -                                   \\
                     &                                           & Assi et al. Label List     & 0.063                                           & 0.48                                          & -                                   \\

                     &                                           & Derived Label List  & 0.170                                           & 1.39                                          & -                                   
\end{tblr}
\end{table}

\label{sec:RQ3_findings}

\textbf{Labels assigned from our derived list are more semantically aligned than those assigned from our baseline label lists.} As shown in Table \ref{tab:big_results_table}, all labeling configurations using our derived list outperform all labeling configurations using any of the three baseline label lists in terms of average cosine similarity. The best-performing baseline label list configuration (Llama using the Catolino et al. label list) achieves 0.143, while even the lowest-performing configuration using our derived list (Llama) achieves 0.147, indicating a consistent improvement in semantic coherence when using our curated label taxonomy.

\textbf{Labels assigned from our derived list are more semantically aligned than the developer-assigned labels.} As shown in Table \ref{tab:big_results_table}, the average cosine similarity of developer-assigned labels is 0.171. Five of the eight configurations using our derived list exceed this value, suggesting that structured, LLM-generated labels match or surpass the originally assigned ones in semantic relevance.

\textbf{The Qwen model without ensembling achieves the highest overall alignment.} As shown in Table \ref{tab:big_results_table}, it achieves an average cosine similarity of 0.178. It is interesting that this labeling configuration achieves a higher cosine similarity relative to all ensembling configurations using our derived list as in contrast, Gemma and Llama's individually produced labels from our derived list are less semantically aligned than any of those generated by the ensembling techniques using our derived list. This discrepancy is likely due to the number of labels assigned per issue report: Gemma and Llama assign an average of 3.32 and 3.72 labels per issue report respectively whereas Qwen assigns an average of 2.31. It is likely that during ensembling, many of the less semantically aligned labels assigned by Gemma and Llama are dropped because they are not assigned by other LLMs which results in a higher average cosine similarity. For Qwen however, it is likely that some of its more semantically aligned labels are not assigned by Gemma and Llama, and thus are excluded in the ensemble, lowering its average cosine similarity relative to Qwen alone. 

% \liam{DIFFERENCE BETWEEN QWEN AND GEMMA + QWEN (2nd highest alignment) IS NOT SIGNIFICANT (p-val = 0.66) (MANN-WHITNEY U-TEST \cite{mann1947test} AS WE DID NOT ASSUME THE 2 POPS FOLLOW A NORMAL DISTRIBUTION}~\shayan{how this would affect our findings? We can discuss in the meeting.} \liam{In my opinion it does not add any relevant information. However, because I cite statistical differences in RQ4, reviewers may find it weird that I don't here so this is what we have to decide}~\shayan{maybe you can say gemma and Qwen and Qwen alone didnt have significant difference therefore picked Qwwn? how about Qwen compare to gemma alone?}

% \usepackage{tabularray}
\begin{table}
\centering
\footnotesize
\caption{Cohen's Kappa interpretation guidelines of Landis and Koch \cite{landis1977measurement}.}
\label{tab:cohens_kappa_interpretation_guidelines}
\begin{tblr}{
  width = \linewidth,
  colspec = {Q[606]Q[304]},
  cells = {c},
  row{1} = {font=\bfseries},
  hlines,
}
Cohen's Kappa Value Range & Interpretation \\
0.00-0.20                 & Slight         \\
0.21-0.40                 & Fair           \\
0.41-0.60                 & Moderate       \\
0.61-0.80                 & Substantial    \\
>0.80                      & Almost Perfect 
\end{tblr}
\end{table}

\textbf{Agreement between label assigner LLMs ranges from Moderate to Almost Perfect across various labeling configurations.} According to the Cohen's Kappa interpretation guidelines of Landis and Koch \cite{landis1977measurement} shown in Table \ref{tab:cohens_kappa_interpretation_guidelines}, labeling configurations using the Colavito et al. label list yield agreement levels from \textit{Substantial} to \textit{Almost Perfect}, whereas labeling configurations using other label lists yield agreement levels from \textit{Moderate} to \textit{Substantial}. It is reasonable that the LLMs using the Colavito et al. label list show higher agreement since they could only choose from four possible labels, whereas the other label lists comprise of at least 9 labels.

\begin{Summary}[ ]{Summary of RQ2}
Constraining LLMs to the taxonomy of 275 labels markedly improves semantic alignment compared to baseline label lists and developer-assigned labels. The \textit{Qwen} model without ensembling techniques achieves the strongest overall alignment, demonstrating that structured label lists enable LLMs to capture nuanced issue themes more effectively while maintaining consistent, interpretable outputs.
\end{Summary}

\subsection{RQ3: Can we enhance our labeling pipeline using retrieval-augmented generation?}
\label{sec:RQ4}

\subsubsection{Motivation} 

While RQ2 showed that constraining LLMs to a coherent label list improves labeling quality, providing the entire list (i.e., 275 labels) within each prompt introduces two key limitations. First, long label lists inflate the context window usage, thereby increasing inference cost and exacerbating positional bias, where models preferentially select labels based on their position in the prompt~\cite{zheng2023judging, pezeshkpour2023large, wang2023primacy}. Second, many labels in the list are irrelevant to a given issue report, which can lead to semantic drift or hallucinated label assignments. To overcome these challenges, we introduce a retrieval-augmented generation (RAG)–based approach that dynamically narrows the label space to a context-specific subset retrieved from a knowledge base of historically labeled issue reports. This design is motivated by the observation that semantically similar issues often share similar labels. By restricting the candidate label set to those used for related issues, the model’s labeling becomes more focused and context aware, leading to an increased labeling accuracy.

\subsubsection{Approach} 
\label{sec:rq4_approach}

\textbf{RAG Knowledge Base Construction.} As discussed in Section \ref{sec:llms_assigning_original_labels}, a foundational component in implementing a RAG-based approach is the RAG knowledge base, which is a database that allows the model to utilize external, domain-specific information during the generation process \cite{gao2023retrieval}. RQ2 yields 33 possible label assignments of issue reports in the training set (see Table \ref{tab:big_results_table}).
% including, for instance, labels assigned by the \textit{Llama} model from the derived label list without ensembling and those generated through unanimous ensembling of \textit{gemma} and \textit{Qwen} using the 4-label approach~\cite{colavito2024leveraging}. \liam{Shayan suggested including the examples, Maram commented them out}
We select the labels assigned by the Qwen model from our derived label list with no ensembling techniques with other LLMs as the foundation for our RAG knowledge base as the labels assigned through this configuration demonstrate the highest semantic alignment with the issue reports in the training set.

\begin{figure*}
        \centering
    \includegraphics[width=0.9\linewidth]{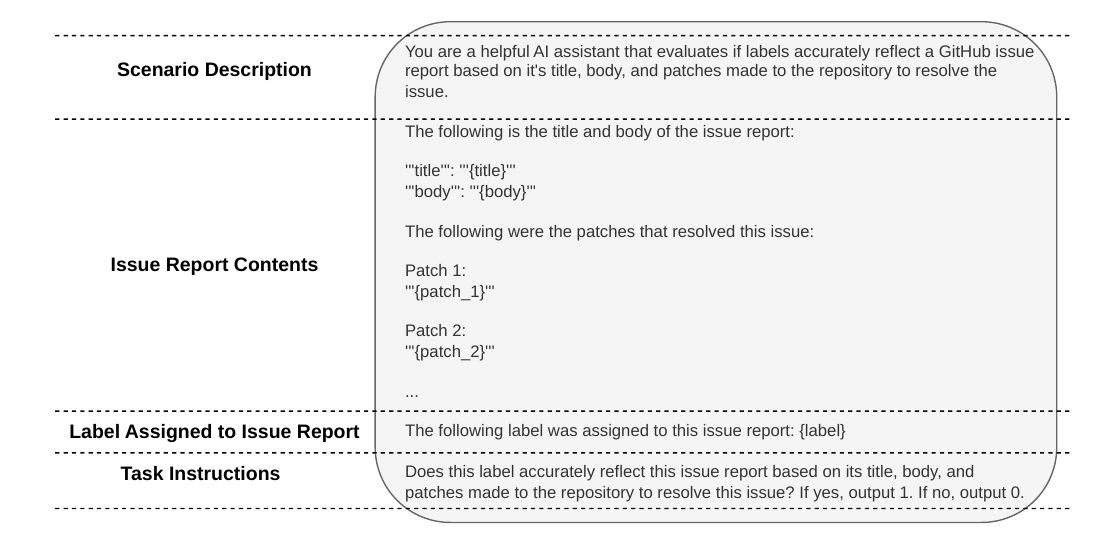}
    \caption{Label evaluation prompt.}
    \label{fig:label_eval_prompt}
\end{figure*}

Next, to ensure accuracy of label assignments, we employ our label evaluator LLM, namely \textit{deepseek-r1:70b}. Specifically, we use the prompt shown in Figure \ref{fig:label_eval_prompt} to assess whether each assigned label ``accurately reflects'' the content of its corresponding issue report based on its title, body, and the associated patch(es) that resolved the issue. The title typically provides a concise summary of the problem \cite{ko2006linguistic}, the body can offer more detailed context (i.e., reproducing steps or test cases, stack traces, and fix suggestions), and the patch(es) reveal the precise functionality that was changed to resolve the issue \cite{weimer2006patches}. Providing all three artifacts allows the evaluator to determine the factual correspondence between an issue and its assigned label. Of the 30,554 labels assigned by Qwen to the 13,210 issue reports in the training set, the evaluator LLM adjudicated that 26,575 (87\%) ``accurately reflected'' their corresponding issue reports. These ``accurate'' labelings form our RAG knowledge base.

Lastly, as with our first motivational study, we embed the historical issue reports (using our text embedding model \textit{all-mpnet-base-v2}) and store these embeddings using the \textit{Faiss} library \cite{douze2024faiss}.

\textbf{Validation of Evaluator LLM via Human Agreement Analysis.} To assess the judgements of our evaluator LLM, the leading author manually reviewed a randomly sampled set of 80 of the 30,554 labels assigned by \textit{Qwen2.5-7B-Instruct} (a statistically representative subset with a 90\% confidence level and 10\% margin of error \cite{hazra2017using}) to the 13,210 issue reports in the training set in the second research question. This author agreed with the evaluator LLM's assessments in 72 of the 80 cases (90\%)\footnote{Evaluations available at \url{https://github.com/24rrvk/LLMIssueLabeling/blob/main/Results_and_Prompts/evaluator_validation/repr_subset_first_review.csv}}. A second author subsequently reviewed the eight disagreements between the leading author and the evaluator LLM and concurred with the evaluator LLM in seven of these cases\footnote{Evaluations available at \url{https://github.com/24rrvk/LLMIssueLabeling/blob/main/Results_and_Prompts/evaluator_validation/repr_subset_second_review.csv}}. Taken together, these results indicate a high level of agreement between human reviewers and the evaluator LLM, providing strong evidence that our evaluation procedure reliably captures label accuracy.

\textbf{Context-Specific Label Assignment.} The first step in the process is to embed the new issue report's title and body to represent their semantic meaning \cite{chersoni2021decoding}. We then retrieve the $k$ issue reports with the most similar embeddings, and therefore most similar semantic meaning, using the \textit{Faiss} library \cite{douze2024faiss} along with their evaluator LLM-validated labels from our knowledge base. The LLM is then prompted to assign the most relevant labels of those retrieved to the new issue report. We experiment with the same RAG-based prompts using the retrieved information introduced in our first motivational study (Section \ref{sec:llms_assigning_original_labels}), namely \textit{Labels Only} and \textit{Labels and Issue Reports}.

\begin{figure*}
        \centering
    \includegraphics[width=1.05\linewidth]{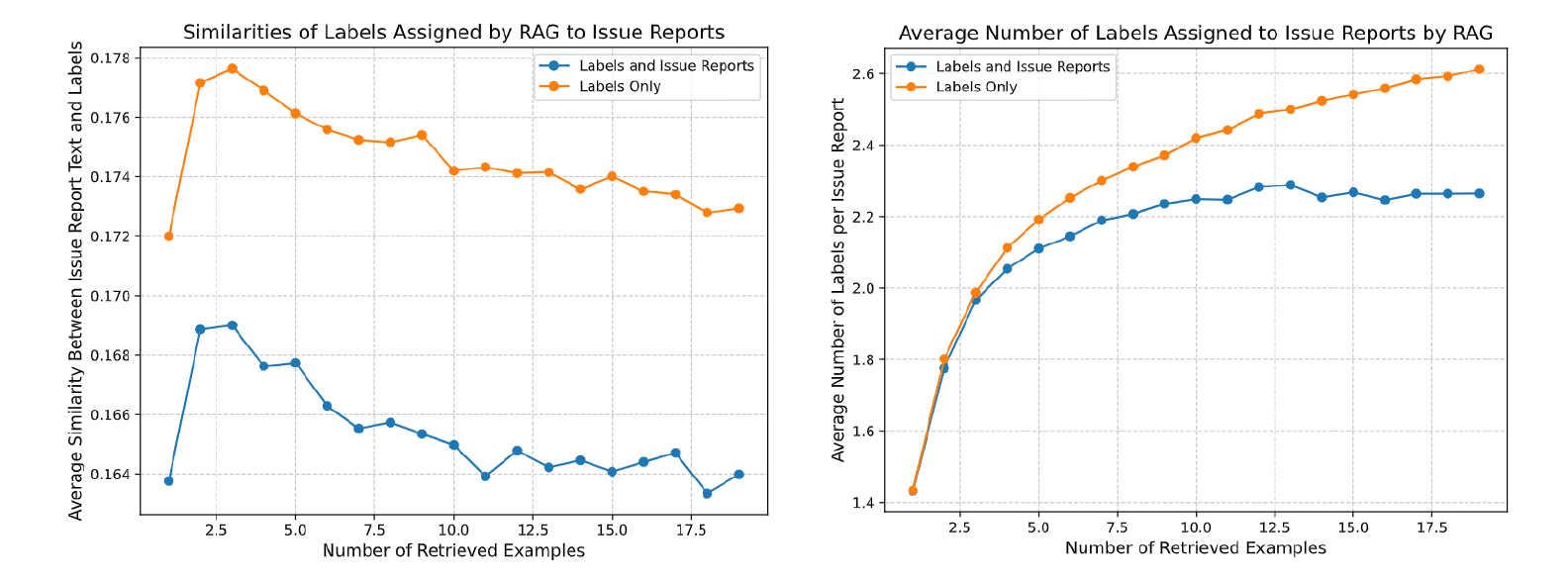}
    \caption{Average cosine similarity and average number of labels per issue reports assigned using the two RAG-based prompts to the 3,210 issue reports in the test set using various numbers of retrieved examples.}
    \label{fig:RAG_results_various_k_vals}
\end{figure*}

\textbf{Context-Specific Label Assignment Evaluation Methodology.} We evaluate both prompts using the information retrieved through our RAG-based approach on the 3,290 issue reports in the test set. For each issue report in the test set, we retrieve the $k$ issue reports in the knowledge base with the most similar embeddings and their validated labels. We then compute the average cosine similarity between the assigned labels and the issue report content (as described in Section \ref{sec:rq3_approach}) across $k=1$ to $k=19$. Label assignments are carried out exclusively with Qwen given that it exhibited the highest average cosine similarity in assigning labels using three of the four label lists, including the derived label list, in RQ3 (see Table \ref{tab:big_results_table}). Figure \ref{fig:RAG_results_various_k_vals} illustrates the results.

As shown in Figure \ref{fig:RAG_results_various_k_vals}, both prompts achieve their highest average cosine similarity at $k=3$ (0.178 for the \textit{labels only} prompt and 0.169 for the \textit{labels and issue reports} prompt). % then the average cosine similarity declines as the value of $k$ increases.  %In terms of the average number of labels assigned per issue report, for the \textit{labels only} prompt, the number of labels assigned continually increases at all values of $k$, whereas this value plateaus for the \textit{labels and issue reports} prompt at $k=11$~\shayan{with the value of?}.
%We then select the highest performing \textit{k} values for each strategy in terms of semantic alignment and 
As a result, we compare the labeling performance of these prompts when $k=3$ against the full 275-label prompt, the three other baselines used in RQ2, i.e., the Colavito et al., Catolino et al., and Assi et al. label lists, and the original developer-assigned labels (excluding the labels not directly referring to the issue itself as in RQ2). Label assignments for all labeling configurations are also exclusively carried out with Qwen for the same reason mentioned above. 

\textbf{Label Assignment Evaluation Metrics.} In addition to semantic alignment measured via cosine similarity, we also report the following three \textbf{evaluator LLM-based metrics}: % ~\shayan{Repeated content. Move to related RQ. In this RQ we just focus on RAG.} \liam{How is this repeated content?? Where was this content mentioned before?? Also, we need to explain this!!!!! This is the setup for the RAG experiments and comparisons against baselines!!!!!!!} 

% \maram{Instead of adding the simple math calculation in the below section, define a formula for each of the metrics. Refer to one of my paper if you need some reference.}

\begin{enumerate}
    \item \textbf{Overall Label Accuracy:} proportion of all labels adjudicated as accurate across all issue reports:
    \begin{equation}
    \text{Overall Label Accuracy} = \frac{\sum_{i=1}^{N} A_i}{\sum_{i=1}^{N} T_i}
    \end{equation}
    where 
    \begin{itemize}
        \item $A_i$ is the number of accurate labels for issue report $i$
        \item $T_i$ is the total number of labels assigned to issue report $i$
        \item $N$ is the total number of issue reports
    \end{itemize}
    \item \textbf{Label Accuracy per Issue Report:} average proportion of accurate labels per issue report:
    \begin{equation}
        \text{Label Accuracy per Issue Report} = \frac{1}{N} \sum_{i=1}^{N} \frac{A_i}{T_i}
    \end{equation}
    \item \textbf{Percentage of Issue Reports with 100\% Label Accuracy:} percentage of issue reports where all assigned labels were accurate:
    \begin{equation}
        \text{Percentage of Issue Reports with 100\% Label Accuracy} = \frac{1}{N} \sum_{i=1}^{N} 1\left( \frac{A_i}{T_i} = 1 \right) \times 100\%
    \end{equation}
    where $1(\cdot)$ is the indicator function which equals 1 if the condition is true and 0 otherwise.
\end{enumerate}

In the previous label assignment studies in this work, namely (1) testing labels assigned by three LLMs from 4 different label lists to 13,210 issue reports in our training set, along with the original labels, and (2) testing labels assigned by both of our RAG-based labeling configurations to the 3,210 issue reports in our test set across number of retrieved example values between 1 and 19 inclusive, the evaluator LLM-based metrics were not used to evaluate labelings. Instead, we used cosine similarity to measure the semantic relatedness between assigned labels and issue report content. This decision was made due to the substantial time cost associated with the evaluator LLM, i.e., it takes an average of 39 seconds per label evaluation. Specifically, each of the two previous label assignment studies require evaluating 171,730 labelings and 121,980 labelings respectively. Given the average of 39 seconds per label evaluation, it would take approximately 77 days to evaluate 171,730 labelings and 55 days to evaluate 121,980 labelings sequentially. 

However, for our final evaluation, we evaluate the labeling performance of our two RAG-based labeling configurations at k = 3 against the full 275-label prompt, the three other baselines used in second research question, i.e., the Colavito et al., Catolino et al, and Assi et al. label lists, and the original developer-assigned labels to the 3,210 issue reports in our test set. Since this means we are only evaluating 22,740 labelings, it is feasible in this context to employ our evaluator LLM to evaluate the accuracy of label assignments. Specifically, given the average of 39 seconds per label evaluation, it would take approximately 10 days to evaluate 22,740 labelings sequentially.

The benefit of using the evaluator LLM in addition to cosine similarity to evaluate label assignments is that while cosine similarity can measure semantic relatedness between assigned labels and issue report content, it cannot directly measure “label correctness”. In contrast, we directly prompt the evaluator LLM to determine whether a label “accurately reflects” an issue report, which acts as a direct assessment of label accuracy. As previously discussed in this section, we also observe high agreement between our evaluator LLM’s judgement and our judgement’s on label accuracy in our \textbf{validation of evaluator LLM via human agreement analysis}. We also compare the performance of labeling configurations according to evaluator LLM-based accuracy metrics and semantic alignment between assigned labels and issue report content measured via cosine similarity to determine whether these metrics yield consistent performance rankings.

To determine if the differences between labeling configurations in terms of average cosine similarity and label accuracy per issue report are statistically significant, we employ the Mann-Whitney U-test \cite{mann1947test}. %\liam{This is where I think it might be weird if there's no mention of this test in RQ3.} 
This non-parametric test compares the distributions of two independent samples and evaluates whether one tends to yield larger values than the other, with the null hypothesis stating there is no difference and the alternative hypothesis stating that there is. We choose this statistical test as the cosine similarity and label accuracy per issue report values for the labeling configurations do not follow normal distributions. % \maram{comment this last sentence, it is not needed} \liam{Shayan agrees that it is important to say WHY you picked the statistical test because otherwise statisticians will ask why you didn't pick a stronger test}

We also calculate the average runtime and the average number of tokens for each prompt, following the same procedure as MS1.

\subsubsection{Findings} 

\begin{table}
\centering
\footnotesize
\caption{Average cosine similarity and label accuracy per label evaluator LLM results for most semantically aligned label assignments per RAG-based prompt against our full derived label list, the Colavito et al., Catolino et al., and Assi et al. label lists, and the original labels assigned to the issue reports in the test set.}
\label{tab:RAG_results}
\begin{tblr}{
  width = \linewidth,
  colspec = {Q[113]Q[129]Q[81]Q[79]Q[112]Q[150]Q[269]},
  row{1} = {c},
  cell{1}{1} = {font=\bfseries},
  cell{1}{2} = {font=\bfseries},
  cell{2}{3} = {c},
  cell{2}{4} = {c},
  cell{2}{5} = {c},
  cell{2}{6} = {c},
  cell{2}{7} = {c},
  cell{3}{1} = {font=\bfseries},
  cell{3}{3} = {c},
  cell{3}{4} = {c},
  cell{3}{5} = {c},
  cell{3}{6} = {c},
  cell{3}{7} = {c},
  cell{4}{1} = {font=\bfseries},
  cell{4}{3} = {c},
  cell{4}{4} = {c},
  cell{4}{5} = {c},
  cell{4}{6} = {c},
  cell{4}{7} = {c},
  cell{5}{1} = {font=\bfseries},
  cell{5}{3} = {c},
  cell{5}{4} = {c},
  cell{5}{5} = {c},
  cell{5}{6} = {c},
  cell{5}{7} = {c},
  cell{6}{1} = {font=\bfseries},
  cell{6}{3} = {c,font=\bfseries},
  cell{6}{4} = {c,font=\bfseries},
  cell{6}{5} = {c},
  cell{6}{6} = {c},
  cell{6}{7} = {c},
  cell{7}{1} = {font=\bfseries},
  cell{7}{3} = {c},
  cell{7}{4} = {c},
  cell{7}{5} = {c,font=\bfseries},
  cell{7}{6} = {c,font=\bfseries},
  cell{7}{7} = {c,font=\bfseries},
  cell{8}{1} = {font=\bfseries},
  cell{8}{3} = {c},
  cell{8}{4} = {c},
  cell{8}{5} = {c},
  cell{8}{6} = {c},
  cell{8}{7} = {c},
  hlines,
}
Label Assigner                          & Label List                                                          & {\textbf{Average}\\\textbf{Cosine}\\\textbf{Similarity}} & {\textbf{Labels}\\\textbf{per Issue}\\\textbf{Report}} & {\textbf{Overall Label}\\\textbf{Accuracy}} & {\textbf{Label Accuracy per}\\\textbf{~Issue Report}} & {\textbf{Percentage of Issue Reports with 100\%}\\\textbf{Label Accuracy}} \\
{\textbf{Developer}} & Project Label List                                                  & 0.176                                                    & 1.89                                                   & 86.77\%                                     & 83.72\%                                               & 73.61\%                                                                    \\
Qwen                                    & {Colavito et al.\\ Label List}                                                    & 0.141                                                    & 1.00                                                   & 77.39\%                                     & 77.39\%                                               & 77.39\%                                                                    \\
Qwen                                    & {Catolino et al.\\ Label List}                                                 & 0.135                                                    & 0.98                                                   &   66.47\%                                           & 66.47\%                                                       &       66.47\%                                                                     \\
Qwen                                    & Assi et al. Label List                                                     & 0.117                                                    & 1.00                                                   &   67.33\%                                          &   67.33\%                                                     &    67.33\%                                                                         \\
Qwen                                    & {Full Derived\\Label List}                                          & 0.180                                                    & 2.36                                                   & 86.65\%                                     & 88.20\%                                               & 73.40\%                                                                    \\
Qwen                                    & {Derived Label\\List, RAG k = 3,\\Labels only\\ prompt}             & 0.178                                                    & 1.99                                                   & 89.84\%                                     & 89.18\%                                               & 81.34\%                                                                    \\
Qwen                                    & {Derived Label\\List, RAG k = 3,\\Labels and issue\\reports prompt} & 0.164                                                    & 1.97                                                   & 84.19\%                                     & 83.07\%                                               & 71.43\%                                                                    
\end{tblr}
\end{table}

\begin{table}
\centering
\footnotesize
\caption{Comparison of average runtime per issue report and average number of tokens per prompt across labeling configurations.}
\label{tab:RQ3_cost_runtime_analysis}
\begin{tblr}{
  width = \linewidth,
  colspec = {Q[125]Q[231]Q[288]Q[294]},
  row{1} = {c,font=\bfseries},
  cell{2}{1} = {font=\bfseries},
  cell{2}{3} = {c},
  cell{2}{4} = {c},
  cell{3}{1} = {font=\bfseries},
  cell{3}{3} = {c},
  cell{3}{4} = {c},
  cell{4}{1} = {font=\bfseries},
  cell{4}{3} = {c},
  cell{4}{4} = {c},
  cell{5}{1} = {font=\bfseries},
  cell{5}{3} = {c},
  cell{5}{4} = {c},
  cell{6}{1} = {font=\bfseries},
  cell{6}{3} = {c},
  cell{6}{4} = {c},
  cell{7}{1} = {font=\bfseries},
  cell{7}{3} = {c},
  cell{7}{4} = {c},
  hlines,
}
Label Assigner & Label List                                                         & Average Runtime per Issue Report (s) & Average Number of Tokens per Prompt \\
Qwen           & Colavito et al. Label List                                                   & \textbf{0.07}                                 & \textbf{338.21}                              \\
Qwen           & Catalino et al. Label List                                                & 0.16                                 & 367.21                              \\
Qwen           & Assi et al. Label List                                                    & 0.14                                 & 379.21                              \\
Qwen           & Full Derived Label List                                            & 0.42                                 & 1,003.21                            \\
Qwen           & {Derived Label List, RAG k = 3, \\Labels only prompt}              & 0.38                                 & 366.70                              \\
Qwen           & {Derived Label List, RAG k = 3, \\Labels and issue reports prompt} & 0.46                                 & 1,354.32                            
\end{tblr}
\end{table}

\textbf{Our RAG-based approach using \textit{Labels only} prompt achieves the highest label accuracy.} As shown in Table \ref{tab:RAG_results}, among all configurations, the RAG \textit{Labels only} approach at $k=3$ achieves the highest performance across all evaluator LLM-based metrics (89.84\% overall label accuracy, 89.18\% label accuracy per issue report, and 81.34\% of issue reports with 100\% label accuracy). While the 275-label prompt attains a higher average cosine similarity, this difference is not statistically significant according to the Mann-Whitney U-test ($p=0.25$). In contrast, the improvement in label accuracy per issue report for the RAG \textit{Labels only} prompt is highly significant according to the Mann-Whitney U-test (p = $1.22\times10^{-10}$). This superior performance likely stems from constraining the LLM to retrieved, validated labels, which encourages more accurate label predictions. Furthermore, as shown in Table \ref{tab:RQ3_cost_runtime_analysis}, the RAG \textit{Labels only} approach at $k=3$ yields efficiency gains over the full 275-label prompt in terms of both average runtime per issue report and average number of tokens per prompt.

\textbf{Labeling configurations with higher average cosine similarity scores tend to achieve higher evaluator LLM-based labeling accuracy scores.} For example, the RAG \textit{Labels only} prompt achieves the highest scores across all evaluator LLM-based metrics (89.84\% overall label accuracy, 89.18\% label accuracy per issue report, and 81.34\% of issue reports with 100\% label accuracy) and the second highest average cosine similarity score (0.178) of all labeling configurations shown in Table \ref{tab:RAG_results}. Similarly, the prompt using the Catolino et al. label list achieves the lowest scores across all evaluator LLM-based metrics (66.47\% overall label accuracy, 66.47\% label accuracy per issue report, and 66.47\% of issue reports with 100\% label accuracy) and the second lowest average cosine similarity score (0.135) of all labeling configurations shown in Table \ref{tab:RAG_results}. Although cosine similarity measures the semantic alignment between assigned labels and issue report content, whereas evaluator LLM-based accuracy directly assesses label correctness, the correspondence between these two metrics suggests that stronger semantic alignment is associated with more accurate label assignments. This indicates that cosine similarity can serve as a useful and scalable indicator of label quality when direct “label correctness” evaluation using an evaluator LLM is computationally impractical.

\textbf{In our RAG-based approach, at the optimal number of retrievals ($k = 3$, 86\% of the retrieved issue reports belong to a repository of the same domain as the repository of their corresponding unseen issue report.} However, even when a retrieved issue report is from a different domain than the corresponding unseen issue report, it often describes a similar type of concern. For example, for issue \#16285 in the GitHub repository \textit{vim/vim}\footnote{\url{https://github.com/vim/vim}}, a text editor, titled ``Finnish menu translation typos and extraneous French block?'', one of the issue reports retrieved at k = 3 is issue \#17752 in the GitHub repository \textit{microsoft/terminal}\footnote{\url{https://github.com/microsoft/terminal}}, a terminal emulator, titled ``Typo in Czech translation of the desktop context menu item''. Although these issue reports are from repositories of different domains, both report natural language translation typos in the software. This example shows that even when the retrieved issues are from a different domain than the corresponding unseen issue report, the similarity-based retrieval mechanism continues to retrieve helpful context by identify issue reports describing comparable underlying problems.

\begin{Summary}[ ]{Summary of RQ3}

The RAG \textit{labels only} prompt achieves the highest labeling accuracy and maintains comparable semantic alignment to the full 275-label prompt while reducing runtime and token usage. Hence, restricting LLMs to retrieved, context-specific labels improves accuracy and efficiency, demonstrating the practical advantage of RAG-based labeling.

\end{Summary}

\subsection{RQ4: How does our coherent label list align with label lists of existing collaborative software repositories?}

\subsubsection{Motivation}

This research question evaluates how the set of 275 labels we derived from 30 diverse software repositories compares with label taxonomies currently used by real-world collaborative software repositories. Through this comparison, we assess whether our synthesized label list provides sufficient coverage to serve as a meaningful baseline or reference point for constructing a new label taxonomy for collaborative software repositories.

\subsubsection{Approach}

We collect label lists from four existing collaborative software repositories. We systematically select two popular GitHub repositories (i.e., more than 87,000 stars) and two less popular GitHub repositories (i.e., less than 99 stars) to evaluate whether our derived label list generalizes across repositories with different levels of popularity. 

As our popular repositories, we select the only two GitHub repositories in the \textit{NLBSE'24 Tool Competition on Issue Report Classification} dataset \cite{kallis2024nlbse} that are not included in our dataset, namely \textit{opencv/opencv}\footnote{\url{https://github.com/opencv/opencv}} and \textit{tensorflow/tensorflow}\footnote{\url{https://github.com/tensorflow/tensorflow}}, which have 87,700 and 195,000 stars respectively. Although \textit{facebook/react}\footnote{\url{https://github.com/facebook/react}} also appears in the NLBSE'24 dataset and is not part of our dataset, we exclude it from this study because the closely related repository \textit{facebook/react-native}\footnote{\url{https://github.com/facebook/react-native}} is already part of our dataset. 

To identify smaller repositories, we query the GitHub REST API for repositories with no more than 99 stars, at least 50 closed issues, and at least 10 labels. We select the first two repositories returned by this query, namely \textit{issp-center-dev/HPhi}\footnote{\url{https://github.com/issp-center-dev/HPhi}} and \textit{sboysel/fedr}\footnote{\url{https://github.com/sboysel/fredr}}. 

We then collect the label lists, collecting 96, 120, 15, and 10 labels from \textit{opencv}, \textit{tensorflow}, \textit{HPhi}, and \textit{fedr} respectively. Next, we pre-process the collected labels by removing those that do not describe the issue itself, following the same procedure used in RQ2 and RQ3. We further remove \textit{project-specific labels}, as the objective of this study is to determine whether our derived label list can serve as a generic baseline applicable across collaborative software repositories. For \textit{opencv}, a computer vision library, project-specific labels primarily correspond to computer vision functionality (e.g., \textit{category: 3d module} and \textit{category: imgproc}). In \textit{tensorflow}, project-specific labels typically refer to individual components or submodules of the library (e.g., \textit{comp:tensorboard} and \textit{ModelOptimizationToolkit}). In contrast, the two smaller repositories do not have project specific labels. These pre-processing steps remove 73, 94, 12, and 4 labels from \textit{opencv}, \textit{tensorflow}, \textit{HPhi}, and \textit{fedr} respectively, leaving 23, 25, 3, and 6 generic labels for these projects. We compare these remaining labels against the 275 labels in our derived label list. The classification of these labels as generic or otherwise can be viewed in our replication package\footnote{\url{https://github.com/24rrvk/LLMIssueLabeling/tree/main/Results_and_Prompts/original_label_lists/projects_not_in_dataset}}.

\textbf{Evaluation Metric.} To quantify how well our derived label list reflects the labels used in existing taxonomies, we measure \textit{coverage} to the extent to which a generic label in the original project label list can be \textit{represented} by at least one sufficiently similar label in our derived list. Intuitively, a label from the original list is considered \textit{covered} if there exists a label in our derived set whose semantic similarity exceeds a pre-defined threshold. We compute \textit{percent coverage} of our derived label list with respect to the generic labels in the original label list using the following formula:

\begin{equation}
\text{Percent Coverage}_{\tau} =
\frac{1}{|O|}
\sum_{o \in O}
\mathbf{1}\!\left(
\max_{l \in L} \mathrm{sim}(o, l) > \tau
\right)
\times 100\%
\end{equation}

where 

\begin{itemize}
    \item $\tau$ is the similarity threshold
    \item $O$ is the set of generic labels in the original label list (in our case the 23, 26, 3, and 6 generic labels in the label lists of \textit{opencv}, \textit{tensorflow}, \textit{HPhi}, and \textit{fedr} respectively)
    \item $L$ is the set of labels in our derived list of 275 labels
    \item sim($o$, $l$) is the similarity between label $o$ in the set of labels $O$ and label $l$ in the set of labels $L$ measured as the cosine similarity between label embeddings generated using our text embedding model \textit{all-mpnet-base-v2}
\end{itemize}

Higher percent coverage values indicate that a larger proportion of labels in the original taxonomy have a semantically similar counterpart in our derived label list, suggesting stronger alignment between the two label sets.

The similarities between all the pre-processed labels in the project label lists and the labels in our label list can be viewed in our replication package\footnote{\url{https://github.com/24rrvk/LLMIssueLabeling/tree/main/Results_and_Prompts/label_similarity_matrices}}. We evaluate percent coverage for $\tau \in {0.1, 0.2, 0.3, 0.4, 0.5, 0.6, 0.7, 0.8, 0.9}$.

\subsubsection{Findings}

\begin{figure*}
        \centering
    \includegraphics[width=0.6\linewidth]{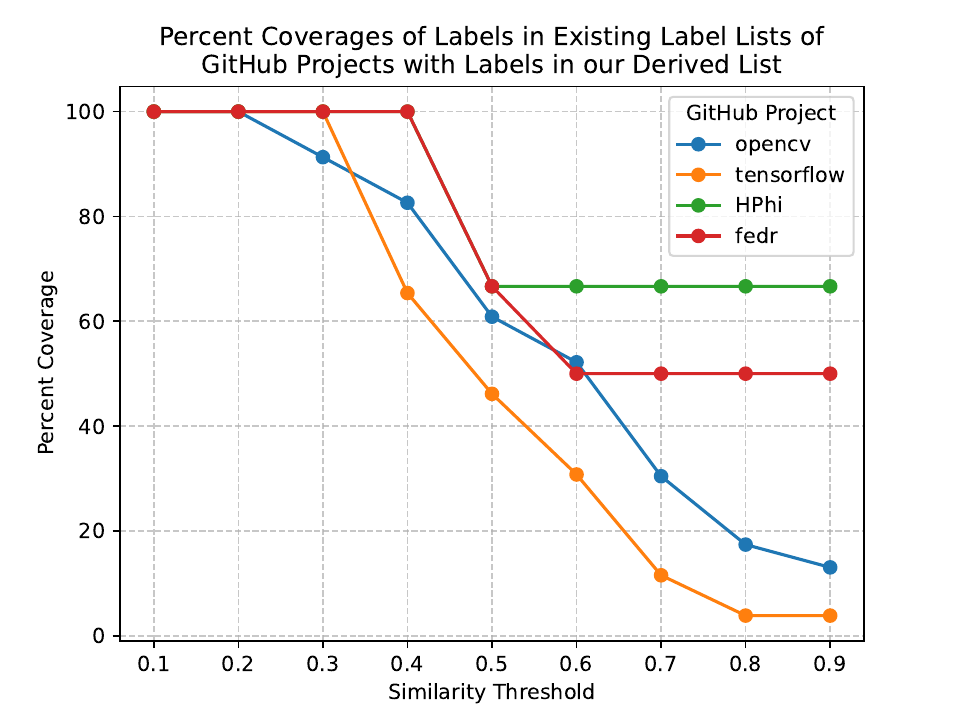}
    \caption{Percentage of labels in label lists of four GitHub projects not in our dataset that have an overlap with at least one label in our derived list of 275 labels at various cosine similarity thresholds.}
    \label{fig:RQ4_results}
\end{figure*}

\textbf{Our derived label list provides substantial coverage of label taxonomies used by existing collaborative software repositories.} As shown in Figure \ref{fig:RQ4_results}, at least 65\% of labels in the original label lists are covered by labels in our derived label list at cosine similarity thresholds of 0.4 or below. This indicates that labels used in these repositories are largely semantically aligned with the generalized issue concepts captured by our derived taxonomy, even under relatively strict similarity requirements. Importantly, a cosine similarity threshold of 0.4 is considered conservative as prior work has shown that a cosine similarity threshold of 0.37 between word vectors corresponds to semantically meaningful similarity \cite{orkphol2019word}. Consequently, achieving substantial coverage at the threshold of 0.4 provides evidence that our derived labels capture genuinely related concepts rather than weak or incidental semantic overlap.

\begin{table}
\centering
\footnotesize
\caption{Labels in used by \textit{fedr}, \textit{HPhi}, \textit{opencv}, and \textit{tensorflow} with a similarity value with a label in our derived list of 275 labels of at least 0.7.}
\label{tab:most_similar_labels}
\begin{tblr}{
  width = \linewidth,
  colspec = {Q[258]Q[273]Q[115]Q[292]},
  row{1} = {c},
  cell{2}{3} = {c},
  cell{3}{3} = {c},
  cell{4}{3} = {c},
  cell{5}{3} = {c},
  cell{6}{3} = {c},
  cell{7}{3} = {c},
  cell{8}{3} = {c},
  cell{9}{3} = {c},
  cell{10}{3} = {c},
  cell{11}{3} = {c},
  cell{12}{3} = {c},
  cell{13}{3} = {c},
  cell{2}{4} = {c},
  cell{3}{4} = {c},
  cell{4}{4} = {c},
  cell{5}{4} = {c},
  cell{6}{4} = {c},
  cell{7}{4} = {c},
  cell{8}{4} = {c},
  cell{9}{4} = {c},
  cell{10}{4} = {c},
  cell{11}{4} = {c},
  cell{12}{4} = {c},
  cell{13}{4} = {c},
  hlines,
}
\textbf{Original Label}    & \textbf{Label in our Derived List} & \textbf{Similarity} & \textbf{Projects of Original Label}              \\
bug                        & bug                                & 1.00                & \textit{opencv}, \textit{fedr}, \textit{HPhi}      \\
cleanup                    & cleanup                            & 1.00                & \textit{opencv}         \\
enhancement                        & enhancement                                & 1.00                & \textit{fedr}, \textit{HPhi}      \\
optimization               & optimization                       & 1.00                & \textit{opencv}         \\
release                    & release                            & 1.00                & \textit{tensorflow} \\
testing                    & testing                            & 1.00                & \textit{fedr} \\
test                       & testing                            & 0.80                & \textit{opencv}         \\
category: documentation    & documentation                      & 0.79                & \textit{opencv}         \\
category: infrastructure   & infrastructure                     & 0.77                & \textit{opencv}         \\
question (invalid tracker) & tracking issue                     & 0.76                & \textit{opencv}         \\
regression issue           & regression                         & 0.75                & \textit{tensorflow} \\
comp:apis                  & api                                & 0.73                & \textit{tensorflow} 
\end{tblr}
\end{table}

\textbf{There are labels in our derived label list that are in label taxonomies used by existing collaborative software repositories.} Table \ref{tab:most_similar_labels} shows labels in the label lists of our four tested GitHub repositories that have a similarity value with a label in our derived list of 275 labels of at least 0.7. It shows that six labels (i.e., \textit{bug}, \textit{cleanup}, \textit{enhancement}, \textit{optimization}, \textit{release}, and \textit{testing}) in our derived label list exactly match labels used at least one of the four repositories. In addition, there are pairs of labels exhibiting near matches such as \textit{test} and \textit{testing} and \textit{regression issue} and \textit{regression}. Other cases differ only in naming conventions, where \textit{opencv} and \textit{tensorflow} prepend categorical prefixes to labels. For example, \textit{opencv} uses the prefix \textit{category: } for their \textit{documentation} and \textit{infrastructure} labels while \textit{tensorflow} uses the prefix \textit{comp:} (short for \textit{component}) to their \textit{apis} label. Overall, these findings indicate that there are labels in our derived list that closely align with labels used in real-world repositories of varying popularity, suggesting that our label list captures commonly accepted concepts despite some differences in naming conventions.

\begin{Summary}[ ]{Summary of RQ4}

Our results show substantial overlap between the label lists of four existing GitHub repositories of varying popularity that were not used in our label generation dataset and our derived list of 275 labels from 30 diverse software repositories. We find up to 100\% coverage of labels in an original list matching a label in our list at similarity thresholds as high as 0.4. This level of overlap, coupled with the fact that there are some labels used by existing repositories that either nearly or exactly match labels in our derived label list, suggests that our derived label list captures widely used issue concepts and provides strong coverage of real-world label taxonomies.

\end{Summary}

\section{Implications}
\label{sec:implications}

In this section, we outline the implications of our study for collaborative software practitioners.

\textbf{Enabling Practical Adoption of Automated Issue Report Labeling.} Despite the recognized value of labeling issue reports, adoption remains low due to the substantial manual effort required to design and maintain suitable label taxonomies and to label new issues \cite{junior2021label, fan2017road}. \textit{LabelMate} directly addresses these barriers by automating both tasks without the requirement of pre-labeled training data or model fine-tuning. This design enables projects of any size to integrate automated labeling into their workflows with minimal setup cost. Collaborative software projects can use our automated labeling techniques in several ways. Projects satisfied with their existing label taxonomy can employ our RAG-based approach: for each incoming issue, semantically similar historically labeled issues can be retrieved and used as contextual evidence, enabling consistent and taxonomy-aligned label assignments with no changes to the existing scheme. Projects seeking to improve their taxonomy or create a new one have two options: (1) adopt our curated list of 275 labels derived from 30 diverse software repositories, offering a broad and empirically grounded taxonomy, or (2) generate a project-specific label list using our taxonomy generation pipeline, ensuring that the resulting labels faithfully reflect the issue landscape of their own repository. Additionally, new types of issues may emerge in future issue reports that are not represented in the historical data used to construct the label taxonomy, such as the development of previously unseen technologies. If developers of a collaborative software project observe that emerging issues are not adequately captured by the existing label taxonomy, they can reapply our label taxonomy generation pipeline to incrementally add these new issue reports to produce an updated taxonomy that reflects the evolving issue space. Alternatively, developers can periodically reapply our pipeline or trigger its reapplication after a predefined number of new issue reports have been submitted to proactively maintain a label taxonomy that remains representative of the project’s issue landscape. 

\textbf{Supporting Automated Issue Resolution.} Software engineering workflows are increasingly incorporating autonomous coding agents to address issue reports \cite{jimenez2024swe}. In this emerging paradigm, issue report labels can play a similar role to that in traditional human-centred triaging: they encode signals that support effective routing. Just as labels help assign issues to developers with appropriate expertise, they can also be used to direct issues to automated coding agents that are best suited for their resolution, enabling more efficient and targeted automated resolution.

\textbf{Real-world Implementation Recommendations.} We provide an example of how our automated labeling pipeline can be implemented in real-world issue triaging workflows in our replication package\footnote{\url{https://github.com/24rrvk/LLMIssueLabeling/tree/main/Practical_Adoption_Pipeline}}. We first provide instructions for loading an LLM on a local machine. We then provide a script for launching a Flask application\footnote{\url{https://flask.palletsprojects.com/en/stable/}} that listens for new issue report submissions to the issue tracking system. The ingestion of new issue reports is implemented using GitHub webhooks, and we also include instructions for configuring a webhook within a GitHub repository. The Flask application subsequently calls a script that pre-processes the issue report text using the steps outlined in Section \ref{sec:issue_report_pre-processing_steps} and assigns labels using the loaded LLM based on the processed text. The LLM-assigned labels can either be directly added to the issue report in the issue tracking system, or a notification can be sent to a project contributor for validation prior to assignment. Even in the latter case, this approach reduces manual effort by shifting the task from selecting appropriate labels from the entire project label taxonomy to simply validating a smaller set of pre-assigned labels, thereby streamlining the labeling process.

% \textbf{Building Trust and Transparency in LLM-Generated Labels.} A major obstacle to integrating LLM-based tools into software engineering workflows is uncertainty over the reliability of their outputs. \textit{LabelMate} addresses this through the inclusion of an evaluator LLM mechanism, where labels generated by a smaller assigner model are validated by a larger evaluator model. This layered design establishes a verifiable feedback process that improves label accuracy and offers project maintainers a transparent means of auditing automated decisions. Such explainable validation supports accountability and can increase developer confidence in adopting LLM-assisted workflows. More broadly, this implication demonstrates how evaluative LLM pipelines can bridge the trust gap that often limits real-world use of generative models in software engineering.

\textbf{Supporting Project-Specific Label Evolution and Insight.} Labels not only categorize issue reports but also capture the evolving structure and priorities of a project. By deriving a coherent taxonomy from historical issue reports, \textit{LabelMate} enables developers to reflect on how their labeling practices evolve over time, thereby revealing recurring issue types, emerging technical themes, or shifts in development focus. This understanding can inform planning decisions such as identifying areas that frequently require fixes, performance improvements, or refactoring. Since the derived taxonomy can be specifically tailored to a given repository, \textit{LabelMate} promotes labeling schemes that can accurately mirror the project's context rather than imposing external standards. As such, the framework can serve as both a practical labeling assistant and an analytical lens for understanding project health and evolution.

\section{Threats to Validity}
\label{sec:threats}

This section outlines potential limitations that may affect the interpretation and generalizablility of our results. We organize these threats according to the standard categories of external, internal, and construct validity.

\subsection{External Validity}

External validity refers to the extent to which the findings of a study generalize beyond its specific experimental setting. Our issue report dataset is constructed from 30 of the 500 most starred GitHub repositories that meet strict inclusion criteria. For example, issue reports were required to have at least one linked resolution patch to be included in our dataset to ensure they corresponded to a meaningful, actionable issue.
% due to the patches' incorporation in label evaluation prompts. 
As such, the labeling performance of our approach observed in this study could be different when evaluating on issue reports from different collaborative software repositories. Such different collaborative software repositories could include proprietary industrial repositories. While we cannot access proprietary industrial repositories, our dataset does include open-source large-scale, industry-developed projects such as \textit{microsoft/vscode}\footnote{\url{https://github.com/microsoft/vscode}}, \textit{microsoft/terminal}\footnote{\url{https://github.com/microsoft/terminal}}, and \textit{facebook/react-native}\footnote{\url{https://github.com/facebook/react-native}}, which can closely resemble many characteristics of industrial software development, including scale, collaborative workflows, and issue management. % \sout{However, the approach can be applied to other issue report dataset to observe the differences in results.} \liam{What's wrong with this? Isn't it good to say that our approach can be applied to different datasets?} 

Additionally, we choose label assigner LLMs that are open-source and relatively lightweight to facilitate efficient and cost-effective label assignment to enable practical adoption of our approach without requiring extensive computational resources. LLMs with larger parameter sizes such as \textit{GPT-5}, \textit{Claude Haiku 4.5}, or \textit{Llama-3.1-70B-Instruct}, may exhibit different behaviours. 

Lastly, in our pipeline that refines candidate labels to a usable taxonomy, we remove candidate labels that were generated by only a single model for a single issue report and candidate labels that refer to the names of specific tools or infrastructure components. While these design decisions may exclude labels that describe specific details of the issue, they support our goal of deriving a coherent list of issue report categories that capture recurring issue types which are applicable across a broader range of collaborative software repositories.

\subsection{Internal Validity}

Internal validity concerns whether the observed effects can be confidently attributed to the design of the experiment rather than to uncontrolled variables. A potential threat to internal validity in this study arises from the sensitivity of LLMs to prompt phrasing as variations in wording or structure can influence model outputs \cite{zheng2023judging, pezeshkpour2023large, wang2023primacy}. To mitigate this, we tested multiple variations of each prompt used in our motivational studies, RQ2, and RQ3 to identify the most stable and effective formulations. Furthermore, we promote transparency by including visual representations of all label generation, assignment, and evaluation prompt templates with justifications for their designs. Despite these precautions, prompt sensitivity remains an inherent characteristic of LLM-based systems and may still introduce variability in performance independent of the experimental design. 

%One potential threat is the reliance on label frequency to select representative labels during clustering~\shayan{myabe rephrase it as we did select based on frequeycy}. This approach may favor broadly applicable labels over more specific but less frequently occurring ones, potentially skewing the final list of labels towards higher semantic alignment by construction~\shayan{what threat it has?}. 

\subsection{Construct Validity}

Construct validity relates to whether the study accurately measures the concepts it is intended to evaluate. In this work, we assess the quality of label assignments using two metrics: (1) semantic alignment, measured as cosine similarity between the embedded representations of labels and issue report content, and (2) an LLM-based evaluation of whether a label ``accurately reflects'' the corresponding issue report. Although these metrics may not provide a complete characterization of label accuracy in all cases, our findings from MS1 demonstrate that the labels originally assigned to issue reports are not reliable ground truth labels. As a result, we consider these measures to be practical and meaningful indicators of label relevance. Moreover, as discussed in Section \ref{sec:rq4_approach}, we observe strong agreement between the evaluator LLM and human assessments of label accuracy, further supporting the validity of this evaluation procedure.

\section{Related Work}
\label{sec:related_work}

This section positions our work within the broader literature on automated issue report labeling and using LLMs to label, annotate, or categorize other software engineering content. 

% We compare our approach to existing methods based on classical machine learning, LLMs, manual labeling techniques, and topic modeling, highlighting key differences in scalability, granularity, and applicability.

\subsection{Three-to-Four Label Approaches}

Kallis et al. \cite{kallis2021predicting} employ a classical natural language processing technique known as \textit{fastText} to label issue reports as either \textit{bug}, \textit{enhancement}, or \textit{question}. Their model is trained and evaluated on issue reports collected from GitHub repositories where one of their labels contained one of the strings \textit{bug}, \textit{enhancement}, or \textit{question}. 

Aracena et al. \cite{aracena2024applying} fine-tune OpenAI's \textit{gpt-3.5-turbo} LLM on the \textit{NLBSE'24 Tool Competition on Issue Report Classification} dataset \cite{kallis2024nlbse}. This dataset is constructed by querying issue reports whose label contain the strings \textit{bug}, \textit{feature}, or \textit{question}, and uses those as the ground truth labels. They fine-tune the model on 1,500 training examples and evaluate it on a separate set of 1,500 test examples.

Colavito et al. \cite{colavito2024leveraging} use the same model and dataset as Aracena et al. However, they express concerns regarding the reliability of contributor-assigned labels. To address this, they select a subset of 400 issue reports and manually label them as either \textit{bug}, \textit{feature}, \textit{question}, or \textit{documentation}. Half of these are used to fine-tune a baseline model using SETFIT, a framework optimized for the fine-tuning of transformer models like Sentence-BERT, while the other half are used for evaluation. The samples used for fine-tuning also serve as few-shot demonstrations in prompts to \textit{gpt-3.5-turbo}.

As noted in Section \ref{sec:intro}, label sets restricted to three or four categories lack the granularity required for practical application. In contrast, our approach enables the derivation of a more detailed set of labels that can also be automatically assigned by an LLM. Furthermore, our approach is more accessible relative to the approaches of Aracena et al. and Colavito et al., as it leverages free open-source LLMs rather than proprietary models that require purchasing tokens through an API. 

\subsection{Assigning Labels used in Practice}

Heo et al. \cite{heo2024comparison} develop \textit{IssueBERT}, a \textit{BERT} model pretrained exclusively on issue reports, and compare its performance against four other BERT variants on two tasks. The first task involves classifying issue reports as either \textit{bug} or \textit{feature}, which, like the aforementioned works, lacks sufficient granularity for real-world use. However, the second task involves assigning any number of labels from the top 10\% most frequently used in the issue report's given repository. While this offers more detailed labeling, the approach relies on project-specific fine-tuning due to varying labeling practices. This requirement limits its applicability to projects that already have an established and consistently used labeling scheme and projects lacking such historical data are unable to leverage this method. Our approach, by contrast, enables these projects to derive and apply a tailored set of relevant labels using an LLM without the pre-requisite of labeled data.

\subsection{Deriving a Label List}
\label{sec:deriving_a_label_list_related_work}

Catolino et al. \cite{catolino2019not} manually evaluate 1,280 bug reports, i.e. they exclude feature requests and other types of issue reports, and derive a set of 9 labels through manual inspection. Arya et al. \cite{arya2019analysis} also manually evaluate 1,326 issue report comments and derive a set of 16 issue report comment labels. Our method provides a more scalable and efficient alternative by using an LLM to automatically generate candidate labels. These candidates are then refined with the help of clustering to group semantically similar labels, allowing users to curate a list of meaningful labels with minimal manual effort.

Assi et al. \cite{assi2023predicting} leverage Embedded Topic Modeling (ETM) \cite{dieng2020topic} to derive 15 categories from a set of 298,548 issue reports. These categories could be used as labels for issue reports. However, using this approach to derive label lists for issue report sets has several limitations: (1) it requires specifying the number of categories, or in this case labels, in advance, (2) the label names must be manually inferred from the set of keywords generated by ETM for each topic, and (3) ETM does not understand the meaning of words in context. For instance, consider the phrases ``this feature is not working properly" and ``you should implement this feature". Clearly, the former phrase is more indicative of a \textit{bug} given that the feature is not working, whereas the latter explicitly states a desire for a new feature to be implement, i.e. a \textit{feature request}. However, ETM may still associate the former phrase with a topic related to feature requests because it relies on word presence rather than contextual meaning. In contrast, our LLM-based approach is context-aware and capable of interpreting nuanced language. It also benefits from external knowledge when generating labels, leading to more accurate labeling of issue reports.

\subsection{LLMs Labeling, Annotating, or Categorizing Software Engineering Content}

LLMs have also been used to label, annotate, or categorize a wide range of software engineering content beyond issue reports. Ahmed et al. \cite{ahmed2025can} test LLMs on 5 software engineering annotation tasks: (1) rating code summarizations on a 4-point Likert scale% \cite{joshi2015likert}
, (2) rating variable name-value inconsistencies on a numerical 5-point scale, (3) determining binarily if a causal relationship exists in natural language software requirement artifacts,  (4) determining whether a pair of functions exhibit the same goals, operations, and effects on a numerical 3-point scale, and (5) determining whether a code change addresses a static analysis warning by assigning the label “closed”, “open”, or “unknown” if the code change removes the warning, if the code change does not remove the warning, or if the warning was deleted or modified in a possibly unrelated way in the code change, making it difficult to confirm whether the warning was actionable, respectively. Ghammam et al. \cite{ghammam2025build} test LLMs on determining which of 22 types of refactoring was conducted based on code changes. Husain et al. \cite{husain2025exploring} evaluate ChatGPT on assigning one of six labels to developer discussions regarding Quantum software engineering. Zhou et al. \cite{zhou2024large} evaluate GPT-3.5 and GPT-4 on the binary classification task on determining if a C or C++ function contains a vulnerability. Cristea et al. \cite{cristea2026malcve} evaluate LLMs on detecting malware in JAR files by classifying them as either \textit{benign}, \textit{suspicious}, or \textit{malicious}. Zeng et al. \cite{zeng2025first} evaluate LLMs on classifying commits as one of ten categories. Chen et al. \cite{chen2026preliminary} evaluate GPT-4 and GPT-5 on classifying the root cause of flaky tests as one of 13 categories. Coutinho et al. \cite{coutinho2026leveraging} evaluate LLMs on classifying the sentiment of a message from a GitHub pull request as either \textit{positive}, \textit{negative}, or \textit{neutral}. These prior works only assign one label to the given content, whereas in our approach, we can assign any number of labels to a given content. Furthermore, these works either provide a pre-defined taxonomy of limited label options \cite{ahmed2025can, chen2026preliminary, coutinho2026leveraging, cristea2026malcve, zeng2025first, zhou2024large} or manually derive the taxonomy of label options \cite{ghammam2025build, husain2025exploring}. Our method provides a more scalable and efficient alternative to manually deriving a label taxonomy that alleviates the manual effort required by leveraging an automated pipeline to curate a taxonomy of labels based on historical issue reports and automatically assign the resulting labels to reports.

\section{Conclusion}
% - I WOULD LIKE TO HAVE A DISCUSSION ABT WHICH METRICS U THINK ARE BEST TO INCLUDE
\label{sec:conclusion}

This study introduces \textit{LabelMate}, a novel end-to-end, domain-adaptive framework for automating the labeling of software issue reports using LLMs. It enables the derivation of a coherent and context-specific label taxonomy from historical issue reports, then employs this taxonomy to label new reports efficiently and accurately. Our findings demonstrate that LLMs, when unconstrained, generate fragmented and redundant label spaces but that these limitations can be overcome through semantic clustering and the use of a fixed, coherent label list. 

By integrating a RAG-based mechanism, \textit{LabelMate} dynamically narrows the candidate label space for each issue report, yielding substantial gains in both label accuracy (up to 89.8\%) and inference efficiency in terms of time and token usage. This makes the framework not only effective but also practical for adoption in real-world repositories of varying sizes and computational capacities. Future work could involve integrating \textit{LabelMate} with real issue triaging workflows and extending this framework toward full triage automation, including duplicate detection and developer assignment, forming a holistic issue report management assistant.

% Future work may explore manual validation of the generation and assignment of labels by collaborative software contributors to assess their human-perceived usefulness. \maram{you need a different future work. This mention of manual weaknes our work} \liam{Isn't this the most important thing though? Trying to figure out if this is actual useful to real people?}
% Beyond automation, \textit{LabelMate} enhances trust and transparency through its evaluator LLM component by enabling explainable validation of generated labels. The derived taxonomies also offer analytical value by reflecting project-specific issue patterns and evolution over time.

% \maram{you can add a sentence saying that in addition, [mention trust] but the way you write sounds as if we do repsonable AI, which we dont}
In summary, \textit{LabelMate} advances the state of automated issue report labeling by enabling scalable, interpretable, and resource-efficient workflows powered by open-source LLMs. 

\bibliographystyle{ACM-Reference-Format}
\bibliography{sample-base}

\end{document}